\documentclass[twocolumn,prd,nofootinbib,superscriptaddress,longbibliography]{revtex4} %

\newcommand\ForInternalReference[1]{}
\newcommand\SkipForEarlyCirculation[1]{}

\newcommand\SkipPP[1]{}
\usepackage[colorlinks=true,citecolor=blue,urlcolor=blue]{hyperref}
\usepackage{verbatim}
\usepackage{graphicx}
\usepackage{dcolumn}
\usepackage{bm}
\usepackage{color}
\usepackage{xcolor}
\usepackage{xspace}
\usepackage{url}
\usepackage{float}
\usepackage{multirow}
\usepackage{adjustbox}
\usepackage[normalem]{ulem}
\usepackage{bm}
\usepackage{placeins}
\usepackage{afterpage}
\usepackage{amsmath,amssymb}
\usepackage{mathtools}
\usepackage{acronym}
\usepackage{pzccal}
\usepackage{physics}
\usepackage{printlen}
\usepackage{subcaption}
\usepackage{array}
\usepackage{pifont}
\DeclareMathAlphabet{\mathcal}{OMS}{cmsy}{m}{n}
\newcommand\optional[1]{}
\acrodef{NR}[NR]{Numerical Relativity}

\usepackage{color}
\definecolor{amber}{rgb}{1.0, 0.75, 0.0}
\definecolor{orange}{rgb}{1.0, 0.5, 0.0}
\definecolor{amaranth}{rgb}{0.9, 0.17, 0.31}

\graphicspath{{./Figures/}}

\def\ltsima{$\; \buildrel < \over \sim \;$}
\def\simlt{\lower.5ex\hbox{\ltsima}}
\def\gtsima{$\; \buildrel > \over \sim \;$}
\def\simgt{\lower.5ex\hbox{\gtsima}}

\newcommand{\UT}{\affiliation{Center for Gravitational Physics, The University of Texas at Austin, Austin, Texas 78712, USA}}
\newcommand{\UCD}{\affiliation{School of Mathematics and Statistics, University College Dublin, Belfield D04 N2E5, Dublin 4, Ireland}}

\begin{document}

\title{Surrogate model of gravitational waveforms for quasi-circular, spin-aligned, asymmetric-mass black hole binaries based on the hybridization of WaSABI-C inspirals with numerical relativity waveforms}

\author{H. L. Iglesias}
\thanks{hector.iglesias@ucd.ie}
\UT \UCD
\author{D. M. Shoemaker}
\UT

\begin{abstract}
With the incorporation of aligned-spin dynamics into self-force waveform models, such as the \texttt{WaSABI-C} inspiral model, and with the release of the SXS collaboration's third public catalog of NR simulations, the gravitational waveform modeling community has made significant advances in the modeling of spinning, asymmetric-mass black hole binaries. In light of these developments, we present \texttt{WaSABINRHybSur2dq15}, the first surrogate model to incorporate information from self-force theory, post-Newtonian theory, and numerical relativity. The model is trained on 65 hybrid waveforms for quasi-circular, spin-aligned, asymmetric-mass black hole binaries with mass ratios spanning 1:1.25 to 1:20 and aligned-spins for the primary black hole ranging -0.5 to 0.5. The hybrid waveforms are generated from a set of \texttt{WaSABI-C} inspirals and NR waveforms from the third SXS public catalog. The model covers the parameter domain $1.25 \leq q \leq 15$ and $|\chi_{1z}| \leq 0.5$, and includes the $(2,2),\, (2,1),\, (3,3),\, (4,4),$ and $(5,5)$ spin-weighted spherical harmonic modes. The model also spans the entire LIGO-Virgo-KAGRA (LVK) frequency band (with $f_{\text{low}}=20$ Hz) for total masses $M \gtrsim 111.5\,M_{\odot}$. We find that the \texttt{WaSABINRHybSur2dq15} model faithfully reproduces the hybrid waveforms with mismatches less than $10^{-3}$ for total masses in the range $80\,M_{\odot} \leq M \leq 300\,M_{\odot}$.
\end{abstract}





\maketitle
\section{Introduction}
\label{sec:intro}

Through the combined efforts of Advanced LIGO \cite{LIGOScientific:2014pky}, Advanced Virgo \cite{VIRGO:2014yos}, and KAGRA \cite{KAGRA:2018plz}, our view of the universe, teeming with compact objects such as black holes (BHs) and neutron stars (NSs), has broadened significantly, owing to the detection of gravitational-wave (GW) transients from binary BHs (BBHs) \cite{LIGOScientific:2016aoc}, binary NSs (BNSs) \cite{LIGOScientific:2017vwq}, and NS-BH (NSBH) binaries \cite{LIGOScientific:2021qlt}, the total of which has now reached 390 \cite{LIGOScientific:2018mvr, LIGOScientific:2021usb, KAGRA:2021vkt, LIGOScientific:2025hdt, LIGOScientific:2025slb, LIGOScientific:2026sit, LIGOScientific:2026wfs}. As the catalog of GW candidates grows with increasing detector sensitivity, the region of binary parameter space for which accurate modeling of GWs becomes essential widens. For example, the detection of GW241011 and GW241110 \cite{LIGOScientific:2025brd}, a pair of rapidly spinning, unequal-mass BH mergers with well-measured spins on the more massive BHs, underscores the need to develop waveform models that are faithful to spinning, asymmetric-mass binaries, which models, when applied to similar GW events, would facilitate our understanding of BH pair formation channels \cite{LIGOScientific:2025pvj, LIGOScientific:2026ctl} and enable us to conduct rigorous tests of general relativity (GR) in regions of strong gravity \cite{LIGOScientific:2016lio, LIGOScientific:2019fpa, LIGOScientific:2018dkp, LIGOScientific:2020tif, LIGOScientific:2021sio, LIGOScientific:2025rid, LIGOScientific:2025wao, LIGOScientific:2026qni, LIGOScientific:2026fcf, LIGOScientific:2026wpt}. Such events will also be of interest when future ground-\cite{Punturo:2010zz, Reitze:2019iox} and space-based \cite{LISA:2017pwj} detectors come online, for the detectors' increased sensitivity at low frequencies will enable observations of the mergers of intermediate-mass BHs $(10^2 \text{--} 10^4\,M_{\odot})$ \cite{LISA:2022yao, ET:2025xjr, Evans:2023euw, Jani:2019ffg, Sesana:2016ljz}.

The modeling of the dynamics of compact object coalescences rests upon multiple approaches \cite{LISAConsortiumWaveformWorkingGroup:2023arg}, ranging from numerical to analytical, each of which has carved out its own regime of validity in the space of binary parameters: numerical relativity (NR), a computational method that directly solves the full, non-linear Einstein field equations, is most accurate in capturing the late inspiral, merger, and ringdown phases of systems with comparable masses \cite{Baumgarte_Shapiro_2010, Pfeiffer:2012pc}; post-Newtonian (PN) theory, a weak-field approximation method, is suitable for describing the early inspiral phase \cite{Blanchet:2013haa}; and gravitational self-force (SF) theory, while best suited for modeling inspirals of systems with extremely disparate mass ratios \cite{Poisson:2011nh, Barack:2018yvs, Pound:2021qin}, has been demonstrated to yield accurate results for close-to-comparable masses \cite{Albertini:2022rfe}, and recent work has shown that the self-force framework can be extended through the transition-to-plunge \cite{Kuchler:2024esj} and merger-ringdown \cite{Kuchler:2025hwx} regimes for non-spinning binaries.

 Following the release of the 1PAT1 inspiral model \cite{Wardell:2021fyy}—a second-order self-force (2SF) waveform model restricted to non-spinning, quasi-circular BBHs at first-post-adiabatic (1PA) order, the order of accuracy required for modeling extreme-mass-ratio inspirals (EMRIs) \cite{LISAConsortiumWaveformWorkingGroup:2023arg}—the self-force community has, over the past year, begun to incorporate (aligned-)spin dynamics into their waveform models. In Ref. \cite{Mathews:2025txc}, for example, the 1PAT1 model was extended to include binaries composed of a slowly spinning primary BH and a secondary BH with generic, precessing spin. Added to this, the authors of Ref. \cite{Honet:2025dho} generalized the next-to-next-to-leading-order transition-to-plunge framework established in their previous work \cite{Kuchler:2024esj} to include the spin of the primary BH. And with the release of the \texttt{WaSABI-C} waveform model \cite{Honet:2025gge, Honet:2025lmk}—a synthesis of first- and second-order SF results with high-order post-Newtonian expansions—which produces spin-(anti)aligned quasi-circular inspirals with dimensionless spin magnitudes up to 0.8, the dynamics of binaries across a wide range of mass ratios, spins, and orbital separations can be accurately captured.

Together with the release of the SXS \cite{SXS} collaboration's third public catalog \cite{Scheel:2025jct} of NR simulations for BBHs, which contains configurations with (anti-)aligned spins and asymmetric\text{--}mass ratios, the emergence of spinning self-force waveform models presents an opportunity to extend the validity of either modeling approach across parameter space—thereby bridging the gap between NR and perturbative methods—through "hybridization", the process by which two waveforms, each generated by a model whose domain of validity overlaps with the other, are sewn together to form an extensive composite waveform. This hybridization procedure enables one to construct inspiral-merger-ringdown (IMR) waveforms that span the full frequency band of GW detectors, which one can leverage by using hybrid waveforms as training data for waveform surrogate models \cite{Field:2013cfa}; such hybrid surrogate models have been investigated in Refs. \cite{Varma:2018mmi, Yoo:2022erv, Yoo:2023spi, Field:2025isp}.



A data-driven approach to waveform modeling, surrogate modeling can be viewed as interpolating waveforms over the binary-parameter space. However, considering that hybrid waveforms are restricted to domains of overlap between different waveform models, a surrogate model trained on such waveforms will inherit this limitation. For instance, several hybrid surrogate models were trained on PN-NR hybrids because the PN mode amplitudes were accurate enough to hybridize with NR simulations for $q \leq 8$ \cite{Varma:2018mmi, Yoo:2023spi}; but, as demonstrated in Fig. 3 of Ref. \cite{Yoo:2022erv}, this method does not work for large mass ratios, for significant deviations between the waveform amplitudes of PN and NR arose at $q=15$ for the $(2,2),\, (2,1),\,(3,3),\,(4,4),$ and $(5,5)$ modes. The \texttt{WaSABI-C} model, however, being a hybrid of PN and SF data itself, does not exhibit such deviations for these modes at $q=15$ (see Sec. \ref{subsec:hybridization} for details). Therefore, in the case of spin-aligned, quasi-circular BBHs, the issue of PN theory being limited to the comparable-mass regime can be overcome by instead using \texttt{WaSABI-C} waveforms in the early inspiral during the hybridization process.



In this work, we build a hybridized NR surrogate model, \texttt{WaSABINRHybSur2dq15}—the first of its kind to incorporate information from SF, PN, and NR approaches—for non-precessing, quasi-circular BBHs with a spinning primary BH and a non-spinning secondary BH. The surrogate model has been trained on 65 hybrid waveforms, each of which consists of a \texttt{WaSABI-C} inspiral waveform that is smoothly stitched to an NR waveform in the late inspiral phase, covering a parameter domain restricted to $1.25 \leq q \leq 15$ and $|\chi_{1z}|\leq 0.5$ with $\chi_{2z}=0$, where $\chi_{1z}$ $(\chi_{2z})$ is the dimensionless spin of the heavier (lighter) BH in the direction of the binary's orbital angular momentum. The duration of each \texttt{WaSABINRHybSur2dq15} waveform is about $30185\,M$, corresponding to a number of GW cycles ranging from 124 to 210, depending on the point in parameter space at which the surrogate is evaluated.


The rest of this paper proceeds as follows. In Section \ref{sec:training_set}, we describe the NR simulations and \texttt{WaSABI-C} inspirals which compose our dataset; and we briefly explain the hybridization method with which we construct our training data. Following this, in Section \ref{sec:Surrogate}, we outline the methodology we used to post-process the training data and build our surrogate waveform model. Next, in Section \ref{sec:error_analysis}, we present an analysis of the errors incurred from the hybridization procedure, from building the surrogate model, and from extrapolating the model to points outside the training parameter space. Lastly, in Section \ref{sec:conclusions}, we conclude with a summary of our results and discuss future work. Throughout this paper, we use geometrized units with $G=1$ and $c=1$. The primary (secondary) BH has mass $m_1$ ($m_2$), dimensionless spin vector $\boldsymbol{\chi}_1=\mathbf{S}_1/(m_1)^2$ ($\boldsymbol{\chi}_2=\mathbf{S}_2/(m_2)^2$), where $m_1 \geq m_2$, and spin magnitude $S_1$ ($S_2$). We define the large mass ratio $q=m_1/m_2 \geq 1$, and the symmetric mass ratio $\nu = m_1 m_2 / M^2$, where $M=m_1 + m_2$ is the total mass. We note that $\nu$ is related to $q$ through $\nu = q / (1+q)^2$. 

\section{Creation of training set}
\label{sec:training_set}
In this section, we describe the steps involved in creating the training set of hybrid waveforms—whose early inspiral portion comes from the \texttt{WaSABI-C} model; late inspiral, merger and ringdown from NR—that will be used to build the new surrogate model \texttt{WaSABINRHybSur2dq15}.

To construct a surrogate model capable of predicting extensive waveforms, we must create a training set of hybrid waveforms and catalog their corresponding binary parameters. Owing to the sparse set of NR simulations in the high\text{--}mass ratio regime $(q \geq 15)$ with (anti-)aligned spins and non-zero secondary spin in the $\hat{z}$-direction (see Sec. \ref{subsec:sxs}), we restrict the parameter space of interest to the 2D region $q \in [1.25, 15]$ and $|\chi_{1z}| \leq 0.5$, with $\chi_{1x}=\chi_{1y}=0$ and $\boldsymbol{\chi_2} = \boldsymbol{0}$; we do not extend the training region to the equal-mass limit as this lies outside the inspiral model's regime of validity (see Sec. \ref{subsec:wasabi}).

In general, we represent the two polarizations of a GW—the plus $(h_+)$ and cross $(h_\times)$ polarizations—as a single complex time-series, $h(t) = h_+ (t) - i\,h_\times (t)$. The complex GW strain can be decomposed into a sum of spin-weighted spherical harmonic modes $h_{\ell m}$ \cite{Newman:1966ub, Goldberg:1966uu} via
\begin{align}
    h(t, \theta, \phi) = \frac{M}{r} \sum_{\ell = 2}^\infty \sum_{m = -\ell}^{\ell} \, h_{\ell m} (t) _{-2}Y_{\ell m} (\theta, \phi),
    \label{eq:strain}
\end{align}
where $_{-2}Y_{\ell m}$ are the spherical harmonics of spin-weight $-2$, $\theta$ and $\phi$ are the polar and azimuthal angles for the direction of GW propagation in the binary's source frame, and $r$ is the distance of the source from the detector. In Eq. (\ref{eq:strain}), the quadrupole modes $(\ell = |m| = 2)$ dominate the sum; nevertheless, as $q$ increases, the subdominant modes (also known as non-quadrupole or higher-order modes) begin to play an important role in parameter estimation of GW sources \cite{Varma:2016dnf, Varma:2014jxa, Capano:2013raa, Shaik:2019dym, Islam:2021zee}. For this reason, our hybrid waveforms include the $(2,2),\,(2,1),\,(3,3),\,(4,4)$ and $(5,5)$ spin-weighted spherical harmonic modes; all other higher-order modes for $\ell \leq 5$ are excluded as explained in Sec. \ref{subsec:hybridization}. We note that the negative $m$ modes do not need to be modeled separately as they can be derived from the positive $m$ modes through the relation $h_{\ell (-m)} = (-1)^\ell h^*_{\ell m}$, where $h^*_{\ell m}$ is the complex conjugate of $h_{\ell m}$, due to the orbital-plane symmetry of non-precessing BBHs.

\subsection{NR Simulations}
\label{subsec:sxs}

The NR waveforms we use to generate the training set for our surrogate model are taken from a set of 65 simulations in the third SXS \cite{SXS} public catalog \cite{Scheel:2025jct} produced with the Spectral Einstein Code (SpEC) \cite{SpEC}. We choose to use SXS simulations because the extensive duration of their waveforms allows us to test the efficacy of our hybridization method (see Sec. \ref{subsec:hybrid_errors}). These simulations are represented as circular markers in Fig. \ref{fig:sxs-dataset}; a catalog of the simulations and their associated metadata can be found in Appendix \ref{appendix:NR}. 

We note that, presently, for $q > 8$, the parameter space of the SXS NR catalog is mostly populated by simulations whose secondary spin $\chi_{2z}$ is zero. For this reason, we choose to restrict our NR dataset to simulations with zero secondary spin. We select the circularly-marked simulations in Fig. \ref{fig:sxs-dataset} because they represent quasi-circular BBHs with (anti-)aligned spins, whose binary parameters lie within the aforesaid training domain: the simulations' mass ratios fall in the range $1.25 \leq q \leq 15$; the maximum dimensionless spin magnitudes for the primary and secondary BHs are $5.0 \times 10^{-1}$ (SXS:BBH:2335) and $1.28 \times 10^{-4}$ (SXS:BBH:1354), respectively; and the largest eccentricity of these simulations is $8.1 \times 10^{-4}$ (SXS:BBH:2484), whereas the median eccentricity is $2.22 \times 10^{-4}$. As of writing, the \texttt{WaSABI-C} model does not handle the $\chi_{1z}=0$ case, requiring inspirals to have primary spin magnitudes $|\chi_{1z}| \geq 10^{-6}$; thus, we place the same restriction on our NR dataset, which results in a minimum value of $2.15 \times 10^{-6}$ (SXS:BBH:2164) for the dimensionless primary spin magnitude.

\begin{figure}
    \centering
    \includegraphics[width=0.95\linewidth]{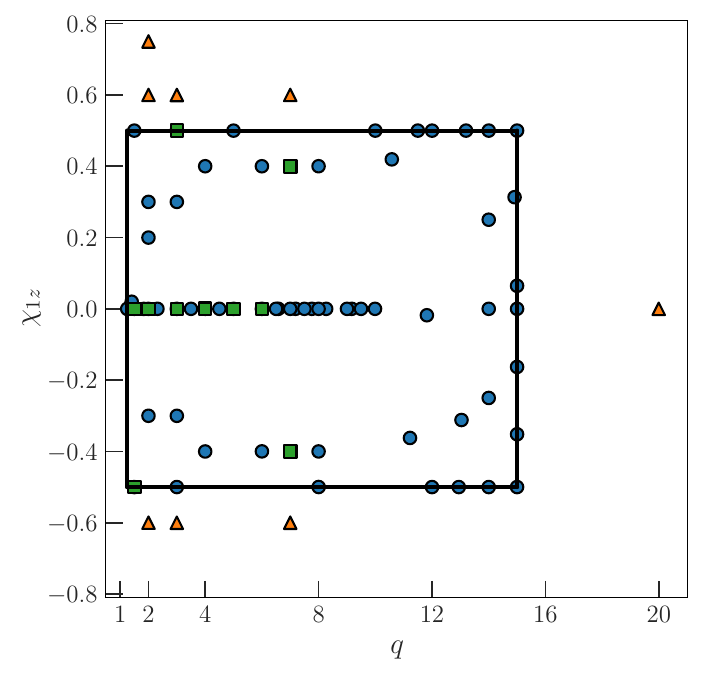}
    \caption{The parameter space containing the 65 NR simulations (circle markers) used to construct the training set for the \texttt{WaSABINRHybSur2dq15} model. We also show 12 simulations (square markers) included in the dataset for which more extensive waveforms exist, and which are used to test hybridization in Sec. \ref{subsec:hybrid_errors}; and 8 simulations (triangle markers) used to test extrapolation in Sec. \ref{subsec:extrapolate}. The boundary of the training set—$1.25 \leq q \leq 15$ and $-0.5 \leq \chi_{1z} \leq 0.5$—is represented by the black rectangle.}
    \label{fig:sxs-dataset}
\end{figure}

For each simulation, the GW strain of Eq. (\ref{eq:strain}) has been extrapolated to future null infinity \cite{Boyle:2009vi}. We use the strain that has been corrected to account for the adventitious motion of the center of mass \cite{Boyle:2015nqa, boyle_2024_12585016}. Additionally, we use the highest level of numerical resolution available for a given simulation. To avoid the impact of junk radiation on the initial masses and spins, we truncate each NR waveform to its reference time, which is provided in the metadata of the corresponding NR simulation \cite{Scheel:2025jct}, and use the values of the masses and spins at the reference time as input parameters for the inspiral model. The start time of these simulations ranges from $1714\,M$ to $18125\,M$ before the peak of the waveform (determined by Eq. (\ref{eq:A_tot})), where $M=m_1+m_2$ is the total Christodoulou mass measured at the reference time. Furthermore, because our hybridization procedure requires higher accuracy during the merger and ringdown portions of these NR waveforms, we use the lowest extrapolation order, $N=2$, following the rule of thumb in Sec. 2.4.1 of Ref. \cite{Boyle:2019kee}. Lastly, we use cubic splines to interpolate the real and imaginary parts of the NR waveform modes with a time step of $\Delta t = 0.1\,M$ to ensure a densely-sampled temporal grid \cite{Virtanen:2019joe}.



\begin{figure*}
    \centering
    \includegraphics[width=1.0\linewidth]{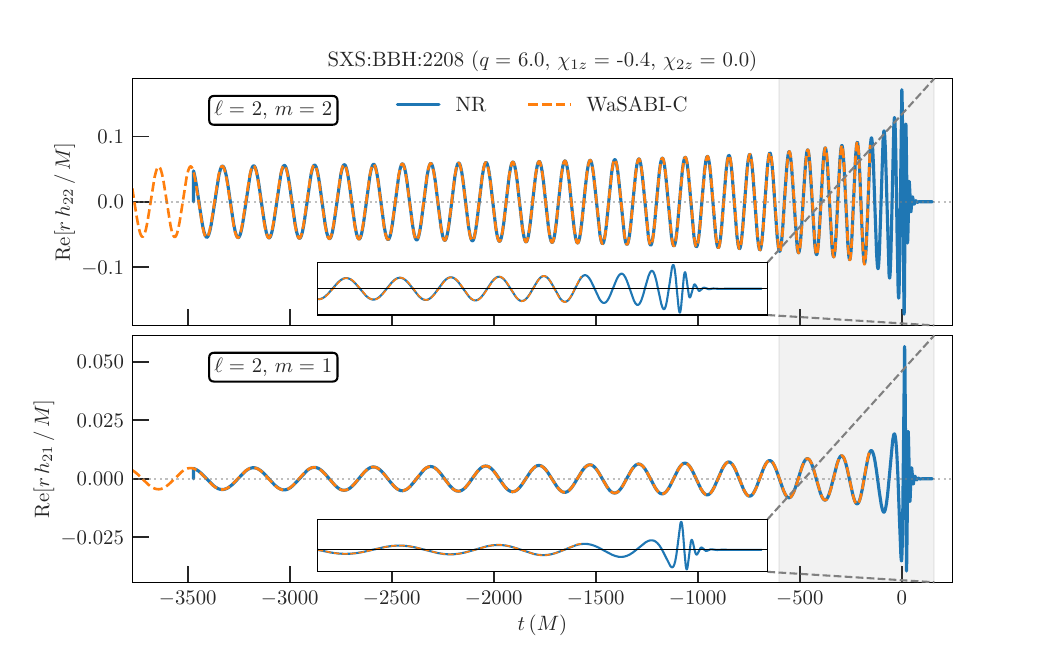}
    \caption{Comparison of NR and \texttt{WaSABI-C} (Sec. \ref{subsec:wasabi}) waveforms for an example simulation. The real part of the $(2,2)$ and $(2,1)$ modes are shown in the top and bottom panels. The binary parameters and SXS identifier of the NR waveform are shown at the top of the plot. The inspiral waveform has been shifted in time and phase using the \texttt{gwtools} package \cite{gwtoolspypi} to align with the NR waveform, whose peak time has been shifted to $t=0$ using Eq. (\ref{eq:A_tot}). The inset shows a close-up of the region near merger.}
    \label{fig:wasabi_NR_comparison}
\end{figure*}

\subsection{WaSABI-C model}
\label{subsec:wasabi}
We generate spin-aligned inspiral waveforms using the \texttt{WaSABI-C} model, a synthesis of SF and PN theory. In this subsection, we provide an overview of the \texttt{WaSABI-C} inspiral model and the settings with which we generated inspirals for hybridization with NR waveforms; for in-depth explanations of the model and its construction, we refer the reader to Refs. \cite{Honet:2025gge, Honet:2025lmk}.

The waveform-generation scheme of the \texttt{WaSABI-C} model employs the multiscale approach developed for spinning, asymmetric-mass binary inspirals \cite{Pound:2021qin, Mathews:2021rod, Mathews:2025nyb}, which divides the scheme into an offline-online split. The offline step consists of solving the Einstein field equations to obtain the GW binding energy and fluxes, in addition to the complex waveform amplitudes, as functions of the binary's mechanical phase-space variables $(m_i, \chi_i, \Omega)$, where $\Omega$ is the orbital frequency; and to express the waveform as a function of the orbital parameters. The subsequent online step, on the other hand, consists of solving a set of differential equations for the binary's parameters, which rapidly produces a trajectory in phase space from which the time dependence of the waveform can be obtained.


As a substitute for the binary parameters $(m_i, \chi_i, \Omega)$, the waveform-generation scheme of the \texttt{WaSABI-C} model utilizes the set $J_a = \{M,\nu,\chi_i,\omega\}$ with $a=\{M,\nu,\chi_i,\omega\}$, where $\omega \approx M\Omega$ is defined as the halved, dimensionless GW frequency of the $(2,2)$ mode, $\omega \equiv M d\psi/dt$, with $\psi \equiv -\frac{1}{2}\,\text{arg}(h_{22})$. The choice to formulate SF expansions in terms of $M$ and $\nu$ instead of the individual masses $m_i$ restores the system's discrete symmetry under the interchange $m_1 \leftrightarrow m_2$ of the two bodies, which improves the accuracy of the expansions in the comparable-mass-ratio regime \cite{Wardell:2021fyy, LeTiec:2014oez, vandeMeent:2020xgc}. Furthermore, as explained in Ref. \cite{Honet:2025gge}, working with waveform frequencies and phases, which are invariant quantities as compared to the orbital frequency, ensures that these variables and functions of these variables represent the same quantity in both SF and PN expansions, clearing the way for a more gauge-invariant hybridization of SF and PN data.

The dynamics of the spin-aligned, quasi-circular inspiral is governed by the evolution equations for $J_a(t)$, which take the form
\begin{align}
    \dv{\psi}{t} = \frac{\omega}{M}, \qquad \dv{J_a}{t} = F_a (J_b),
    \label{eq:evolution_eqns}
\end{align}
with initial conditions $\psi (t_0) = \psi_0$ and $J_a (t_0) = J_{0a}$. Using flux-balance laws \cite{Teukolsky:1974yv, Thorne:1984mz, Tagoshi:1997jy, Alvi:2001mx, Poisson:2004cw, Ashtekar:2004cn}, the forcing functions can be written in terms of the binding energy $E^{\text{bind}} \coloneq M_{\text{B}} - m_1 - m_2$, where $M_{\text{B}}$ is the Bondi mass; the GW energy flux to infinity $\mathcal{F}_{\infty} = -dM_{\text{B}} / dt$, using the Bondi-Sachs mass-loss formula \cite{10.1098/rspa.1962.0161, 10.1098/rspa.1962.0206}; and the GW energy and momentum fluxes through the $i$th BH's horizon, $\mathcal{F}_{\mathcal{H}_i} = dm_i / dt$ and $\mathcal{G}_{\mathcal{H}_i} = dS_i / dt$, respectively.

To overcome the obstacle presented by the lack of second-order SF results for spinning primary BHs, the functions  $E^{\text{bind}}$, $\mathcal{F}_{\infty}$, $\mathcal{F}_{\mathcal{H}_i}$, and $\mathcal{G}_{\mathcal{H}_i}$, along with the complex waveform amplitudes $A_{\ell m}$, are determined from a hybridization of SF and PN data at the highest available order in each expansion. For any invariant function $f$, whose SF expansion $f^{\text{SF}}$ is a power series in $\nu$, and PN expansion a series in $x \equiv \omega^{2/3}$, an SF+PN composite solution can be constructed by adding together the SF and PN expansions and subtracting the common terms $f^{\text{SF}|\text{PN}}$,
\begin{align}
    f^{\text{SF} + \text{PN}} = f^{\text{SF}} + f^{\text{PN}} - f^{\text{SF}|\text{PN}},
    \label{eq:composite}
\end{align}
which cancels the doubly-counted terms, as is standard in the method of matched asymptotic expansions \cite{alma991035990530507336}. After applying this hybridization procedure to the GW energy, fluxes, and complex amplitudes, the forcing functions are then given in terms of the composite quantities. We note that the forcing functions in the right-hand side of Eq. (\ref{eq:evolution_eqns}) are written exactly and are not re-expanded in either SF or PN limits; this choice has been shown in Refs. \cite{Burke:2023lno, Mathews:2025txc} to lead to a notable improvement in the accuracy of self-force models for comparable masses.

By inserting the hybrid quantities in the forcing functions and then integrating Eq. (\ref{eq:evolution_eqns}) given initial conditions for the binary mechanical parameters, we obtain an inspiral waveform which benefits from the weak-field and comparable-mass validity of PN theory and the strong-field and asymmetric-mass validity of SF theory. According to Ref. \cite{Honet:2025lmk}, \texttt{WaSABI-C} includes second-order SF flux results for $\chi_{1}=0$ and linear-in-$\chi_2$ results for generic $\chi_1$; and all available PN terms, excluding quartic-in-spin contributions; details on the specific PN and SF data used to construct the model can be found in Refs. \cite{Mathews:2025txc, Honet:2025gge}.

The \texttt{WaSABI-C} model is available in the Mathematica package \href{https://bhptoolkit.org/WaSABI/}{\textrm{W\footnotesize A\normalsize SABI}} \cite{wardell_2025_17405583} from the Black Hole Perturbation Toolkit \cite{BHPToolkit}. For this work, we use version 1.1.1. The model uses the following binary parameters to generate inspiral waveforms: the total mass $M$, the symmetric mass ratio $\nu$, the dimensionless spins $\chi_{1z}$ and $\chi_{2z}$, and the initial orbital phase and frequency. Given that each NR waveform is truncated at its reference time (see Sec. \ref{subsec:sxs}), the Christodoulou masses $m_1$ and $m_2$, as well as the dimensionless spins $\chi_{1z}$ and $\chi_{2z}$, measured at reference time are used to define the simulation's mass and spin parameters; thus, for consistency, and to achieve a meaningful comparison between the inspirals and NR waveforms, we use these same masses and spins to define $M$, $\nu$, $\chi_{1z}$, and $\chi_{2z}$ in the \texttt{WaSABI-C} model. Also, we set the initial phase to zero and the initial frequency $\omega_0$ to $8 \times 10^{-3}$, a value smaller than all the orbital frequencies of the NR waveforms measured at reference time (see Table \ref{table:sim_list}). Lastly, the waveform modes are generated up to $\ell=5$ with a time step of $0.5\,M$, after which we interpolate the modes using a cubic-spline interpolation scheme with a uniform time step of $\Delta\,t = 0.1\,M$ \cite{Virtanen:2019joe}. We note that the \texttt{WaSABI-C} inspiral terminates at the point the forcing function $F_{\omega}$, which drives the evolution of $\omega$ in Eq. (\ref{eq:evolution_eqns}), reaches its maximum value.

Figure \ref{fig:wasabi_NR_comparison} shows an example of an NR waveform with binary parameters $(q,\chi_{1z},\chi_{2z})=(6,-0.4,0)$ and its corresponding \texttt{WaSABI-C} inspiral waveform. Using the \texttt{gwtools} package \cite{gwtoolspypi}, we apply the optimal shifts in time and phase to the inspiral waveform that align it with the NR waveform. We find that there is good agreement between the waveforms in the early inspiral, and little to no dephasing in the inspiral waveform's last few cycles before termination.

\begin{figure}
    \centering 
    \makebox[\linewidth][c]{ \includegraphics[width=1.05\linewidth]{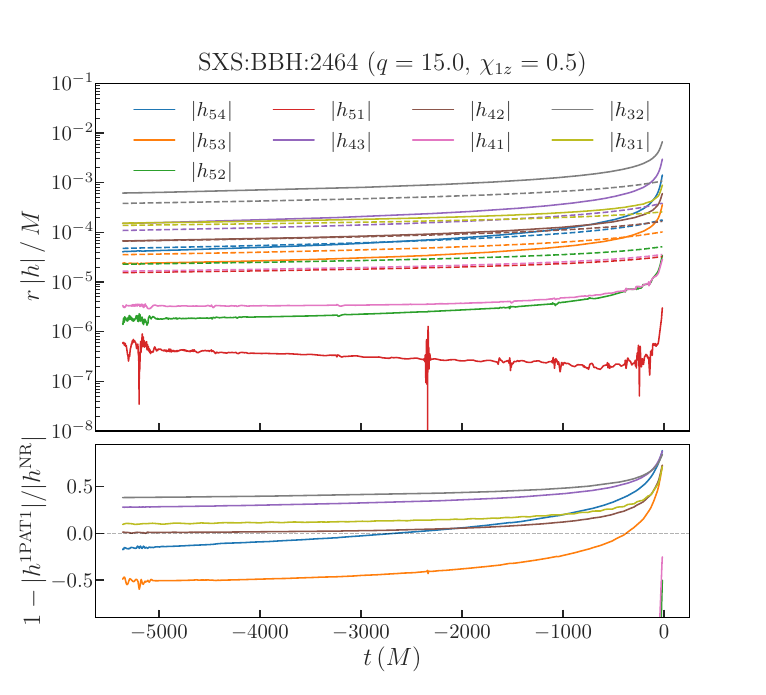}}
    \caption{\textit{Top}: Waveform amplitudes for \texttt{WaSABI-C} (dashed) and NR (solid) for a set of spherical harmonic modes. The binary parameters and SXS identifier of the NR waveform are shown at the top of the plot. The waveforms have been aligned using the \texttt{gwtools} package; and their peak times have been shifted to $t=0$. We show the modes that have been excluded from the hybridization procedure. \textit{Bottom}: Relative error in the waveform amplitudes for the excluded modes.}
    \label{fig:excluded-modes}
\end{figure}

\begin{figure}
    \centering 
    \makebox[\linewidth][c]{ \includegraphics[width=1.05\linewidth]{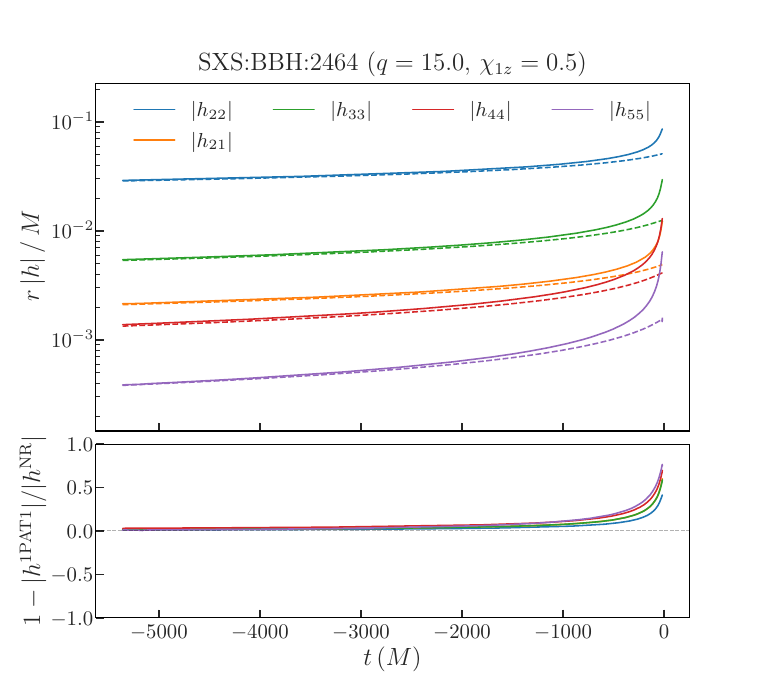}}
    \caption{\textit{Top}: Waveform amplitudes for \texttt{WaSABI-C} (dashed) and NR (solid) for a set of spherical harmonic modes. The binary parameters and SXS identifier of the NR waveform are shown at the top of the plot. The waveforms have been aligned using the \texttt{gwtools} package; and their peak times have been shifted to $t=0$. We show the modes that are included in the hybridization procedure. \textit{Bottom}: Relative error in the waveform amplitudes for the included modes.}
    \label{fig:included-modes}
\end{figure}


\begin{figure*}
    \centering
    \makebox[\textwidth][c]{\includegraphics[width=1.1\linewidth]{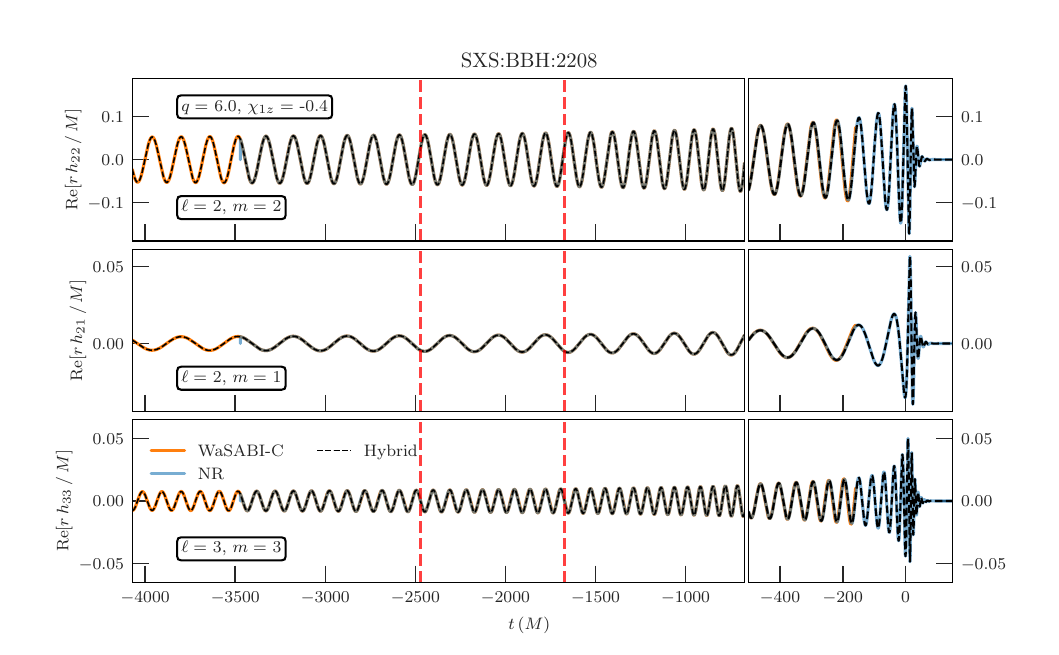}}
    \caption{An example hybrid waveform for a simulation with binary parameters $(q,\,\chi_{1z})=(6,\, -0.4)$. We show the $(2,2)$, $(2,1)$, and $(3,3)$ modes in the top, middle, and bottom rows of the plot. The SXS identifier of the NR waveform is shown at the top of the plot. The vertical dashed red lines represent the matching window. Portions of the waveforms that appear gray are where the \texttt{WaSABI-C} model and NR become indistinguishable. We note that the plot shows the \texttt{WaSABI-C} and NR waveforms after having been transformed into the hybridization frame, and that the time has been shifted such that the peak of the total amplitude occurs at $t=0$.}
    \label{fig:hybrid-modes-1}
\end{figure*}

\begin{figure*}
    \centering
    \makebox[\textwidth][c]{\includegraphics[width=1.1\linewidth]{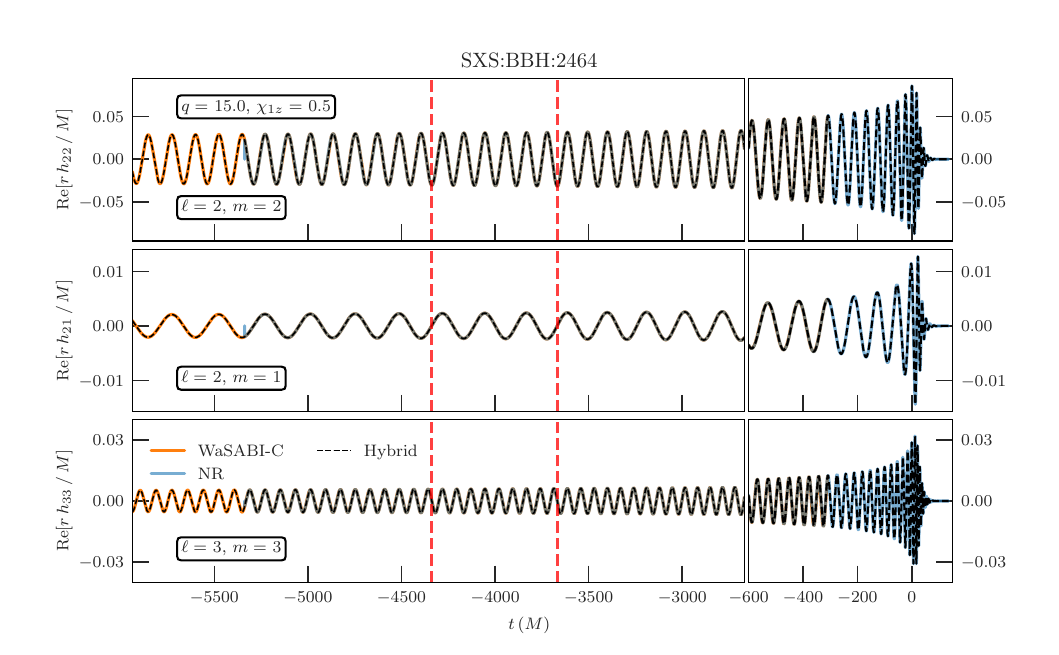}}
    \caption{Same as Fig. \ref{fig:hybrid-modes-1}, but now for a simulation with binary parameters $(q,\,\chi_{1z})=(15,\, 0.5)$.}
    \label{fig:hybrid-modes-2}
\end{figure*}

\subsection{Hybridization}
\label{subsec:hybridization}
To hybridize NR and SF+PN waveforms describing
quasi-circular BBHs with spin-aligned primary BHs, we follow the hybridization procedure outlined in Sec. V of Ref. \cite{Varma:2018mmi}. In contrast with our previous paper \cite{26wy-s7tf}, we are following Ref. \cite{Varma:2018mmi} in our choice of the matching window, setting it at $1000\,M$ after the NR reference time, when constructing the hybrid waveforms. We note that we use a three-orbit long window, whose duration in time is determined by using the phase of the $(2,2)$ mode of the NR waveform.

Though the \texttt{WaSABI-C} model can generate modes up to $\ell=5$, we have chosen to exclude the $(3,2)$, $(3,1)$, $(4,3)$, $(4,2)$, $(4,1)$, $(5,4)$, $(5,3)$, $(5,2)$, and $(5,1)$ modes, for the waveform amplitudes of said modes from the \texttt{WaSABI-C} model do not, in general, agree with those of NR, as demonstrated in Fig. \ref{fig:excluded-modes}. For instance, we find that the $(5,1)$ mode exhibits the largest relative error in the waveform amplitude, with the $(5,2)$ and $(4,1)$ modes also showing significantly large relative errors. Furthermore, while we observe that, of the remaining excluded modes, the $(3,2)$ mode amplitude tends to exhibit the smallest relative error; nevertheless, we find that this trend is not consistent across points in our dataset; consequently, the inclusion of the $(3,2)$ mode in the list of available modes—namely, the $(2,2)$, $(2,1)$, $(3,3)$, $(4,4)$, and $(5,5)$ modes (see Fig. \ref{fig:included-modes})—will bias the hybrid training data.

After creating the hybrid waveforms, we use cubic splines to interpolate the real and imaginary parts of each hybrid waveform mode with a uniform time step of $\Delta\,t = 0.1\,M$ to ensure a densely-sampled temporal grid \cite{Virtanen:2019joe}. We present in Fig. \ref{fig:hybrid-modes-1} an example of the hybrid, NR, and \texttt{WaSABI-C} waveforms at the binary parameter point $(q,\, \chi_{1z}) = (6,\, -0.4)$; likewise in Fig. \ref{fig:hybrid-modes-2} for $(q,\, \chi_{1z}) = (15,\, 0.5)$.

\section{Surrogate model}
\label{sec:Surrogate}
In this section, we describe our method to construct a surrogate model of gravitational waveforms trained on the dataset of hybrid waveforms produced in Sec. \ref{subsec:hybridization}. We follow the methodology developed in Refs. \cite{Varma:2018mmi, Yoo:2022erv, Yoo:2023spi}, deviating only in our choice of data decomposition.

\subsection{Post-processing the training data}
\label{subsec:postprocess}
\subsubsection{Temporal alignment}
\label{subsubsec:time_align}
In order to construct smooth parametric fits of the waveform data pieces that will compose the surrogate model, we must first align all the waveforms in our training dataset such that the peak of their total amplitude,
\begin{align}
    A_{\text{tot}}(t) = \sqrt{\sum_{\ell m} |h_{\ell m} (t)|^2},
    \label{eq:A_tot}
\end{align}
is at its maximum value at the same time. To do this alignment, we apply a time shift to each waveform such that their peak time, obtained from a quadratic fit using the nearest five points to the discrete maximum of $A_{\text{tot}}(t)$, is located at $t=0$. In calculating the sum in Eq. (\ref{eq:A_tot}), we use all the available modes of the hybrid waveform, including the $m < 0$ modes.

\subsubsection{Common time array and downsampling}
\label{subsubsec:downsample}
Next, we ensure that the training waveforms share a common temporal grid, a requirement of the surrogate-building procedure. The length of each hybrid waveform is set by one's choice of the \texttt{WaSABI-C} inspiral's initial orbital frequency $\omega_{\text{orb}}$, which can be approximated via $\omega_{\text{orb}} \approx \frac{1}{2} \dv{\phi_{22}}{t}$, where $\phi_{22}$ is the phase of the $(2,2)$ mode; with the settings we've chosen in Sec. \ref{subsec:wasabi}, this corresponds to a starting frequency of $\omega_{\text{orb}} = 8 \times 10^{-3} \, \text{rad}/M$ for all waveforms. Although the waveforms have the same initial orbital frequency, they will, in general, have different durations owing to differences in their intrinsic parameters—here, mass ratio and primary spin. For this reason, we truncate each waveform such that their starting time matches that of the shortest waveform, which we find to be $t = -30070\,M$; additionally, we retain times up to $t=115\,M$, which is sufficient to capture the entire ringdown, resulting in a common time array in which $t \in [-30070\,M, 115\,M]$.

After the truncation, the largest starting orbital frequency in the training set is $\omega_{\text{orb}} = 1.4 \times 10^{-2}\,\text{rad}/M$, which establishes the low-frequency limit of validity of the surrogate model. Assuming that 20 Hz is the lowest frequency that can be measured by the LIGO-Virgo-KAGRA (LVK) observatories, the $(2,2)$ mode of the surrogate model is valid for total masses $M \geq 44.6\,M_{\odot}$. Since the $(5,5)$ mode is the highest subdominant mode available in the model, whose frequency is 5/2 times that of the (2,2) mode, all modes of the surrogate model are then valid for total masses $M \gtrsim 111.5\,M_{\odot}$.

With a constant time step of $0.1\,M$, these waveforms, lasting hundreds of thousands of $M$, prove too long to work with as regards the surrogate-building procedure owing to computational costs, leading us to downsample them to reduce said costs. We do so by retaining five time samples per orbit for the shortest hybrid waveform for times $t < -1000\,M$—that is to say, by keeping one sample every $\pi$ radians prior to $t = -1000\,M$ with respect to $\phi_{22}$; for $t \geq -1000\,M$, we use a uniform step size of $0.1\,M$, ensuring that we cover the high-frequency portion of the waveform—the late inspiral and merger-ringdown phases—with a denser sampling rate; finally, we use a cubic spline to interpolate each waveform mode onto the downsampled common time array.

\subsubsection{Phase alignment}
\label{subsubsec:phase_align}
After downsampling and interpolating to a common temporal grid, we align the waveforms' phases by rotating each waveform about the $\hat{z}$-axis such that at $t=-1000\,M$ the phase of the $(2,2)$ mode is set to zero. This rotation ensures that each waveform corresponds to a black hole binary in which the heavier black hole is located on the $\hat{x}$-axis. We note, however, that this phase alignment fixes the physical rotation up to a phase shift of $\pi$, which ambiguity is fixed by requiring that the phase of the $(2,1)$ mode lies between $[0,\pi)$. 



\subsubsection{Data decomposition}
\label{subsubsec:data_decomp}
Rather than constructing a surrogate model of the inertial frame strain $h_{\ell m}$, we decompose the rapidly oscillating waveform modes into ``waveform data pieces'', slowly varying functions of time that are easier to model, and create a separate surrogate model for each data piece. For non-precessing, quasi-circular waveforms, there are two approaches to decomposing the waveform data one can take: (i) transform the waveforms into the co-orbital frame—as in Refs. \cite{Varma:2018mmi, Yoo:2022erv, Yoo:2023spi, Islam:2022laz}—and extract the amplitude $A_{22}$ and phase $\phi_{22}$ of the $(2,2)$ mode, as well as the real and imaginary parts of the co-orbital frame strain $h_{\ell m}^C$ for all other modes; or (ii) decompose each mode of the complex strain $h_{\ell m}$ into an amplitude $A_{\ell m}$ and phase $\phi_{\ell m}$, as in Ref. \cite{Rink:2024swg},
\begin{align}
    h_{\ell m} = A_{\ell m}\,e^{-i\,\phi_{\ell m}}.
    \label{eq:hlm}
\end{align}
The advantage to defining the data pieces in the co-orbital frame lies in its avoiding the divergent behavior of the phase of the odd $m$-modes in the equal-mass limit. Yet, as we do not model equal-mass BBHs, owing to the inspiral model's regime of validity, our phases are well-defined throughout the training parameter space, and so the waveform data pieces we choose to model are $A_{\ell m}$ and $\phi_{\ell m}$.

\subsection{Constructing the surrogate}
\label{subsec:constructing_surrogate}


With the waveform data pieces in hand, we construct a surrogate model for each data piece following the procedure described in Ref. \cite{Varma:2018mmi}, which we summarize below, after which we combine the surrogates to build a gravitational waveform surrogate model. According to Ref. \cite{Field:2013cfa}, to construct a surrogate model, whereby one can predict a gravitational waveform at an arbitrary parameter $\boldsymbol{\lambda}$, one must build a reduced basis for each waveform data piece and evaluate the data pieces at algorithmically chosen times. Whereas the training set is well approximated by an expansion of the reduced basis, it is the combination of this basis and the evaluations at the select times which enable one to approximate waveforms outside the training set but within a compact parameter domain.

We begin by building a reduced basis for each waveform data piece. We use a greedy algorithm \cite{Field:2013cfa, Field:2011mf, Blackman:2017dfb}—described in Sec. III A and Appendix A of Ref. \cite{Field:2013cfa}, as well as Sec. II E of Ref. \cite{Yoo:2023spi}—which selects the most representative elements of the data piece through an iterative procedure. The greedy algorithm first selects the element with the largest $L^2$-norm, orthonormalizes the element, and adds the result to the basis set as the first basis function. Then, until a pre-determined number of basis functions is obtained, the algorithm computes the projection error between each element of the training data and the basis set (excluding elements added to the basis set), selects the element with the largest error, orthonormalizes the element against the current basis set, and adds the result to the basis set. We determine the number of basis functions used to model each data piece through trial and error; that is to say, we increase the number of basis functions until the addition of another basis function ceases to aid in the minimization of the projection errors or introduces noise. The number of basis functions that is used for each waveform data piece is shown in Table \ref{table:basis_functions}. When computing the projection errors, we only use data up to $50\,M$ after the peak, as we find this helps avoid noisy basis functions.





In the next step, we employ the Empirical Interpolant Method (EIM) \cite{CRMATH_2004__339_9_667_0, Yvon_Maday, refId0}, a framework which, using a greedy procedure described in Sec. III B and Appendix B of Ref. \cite{Field:2013cfa}, allows us to identify a sparse subset of time samples—called empirical times or nodes—from which we can build an empirical interpolant (i.e. surrogate model) that can accurately reconstruct a fiducial waveform data piece at any other time. By construction, the empirical times are uniquely defined by the greedy basis functions—a greedy basis of aligned-spin waveforms, for example, will differ in its distribution of the empirical nodes when compared to a basis of precessing waveforms; and the number of nodes is the same as the number of basis functions. Following the methodology of Ref. \cite{Varma:2018mmi}, we require that the starting sample of a waveform data piece is always included as an empirical node; this provides each interpolant with an ``anchor point" that ensures that its corresponding data piece starts with the correct value. Moreover, we restrict the times from which the nodes are selected to $t < 50\,M$ to ensure that little to no numerical noise is being modeled by the interpolants, particularly for the phase data pieces. To select the empirical nodes, we use the Python package \texttt{RomPy} \cite{Field:2013cfa, rompypypi}.


\begin{figure*}
    \centering
    \makebox[\textwidth][c]{\includegraphics[width=1.1\linewidth]{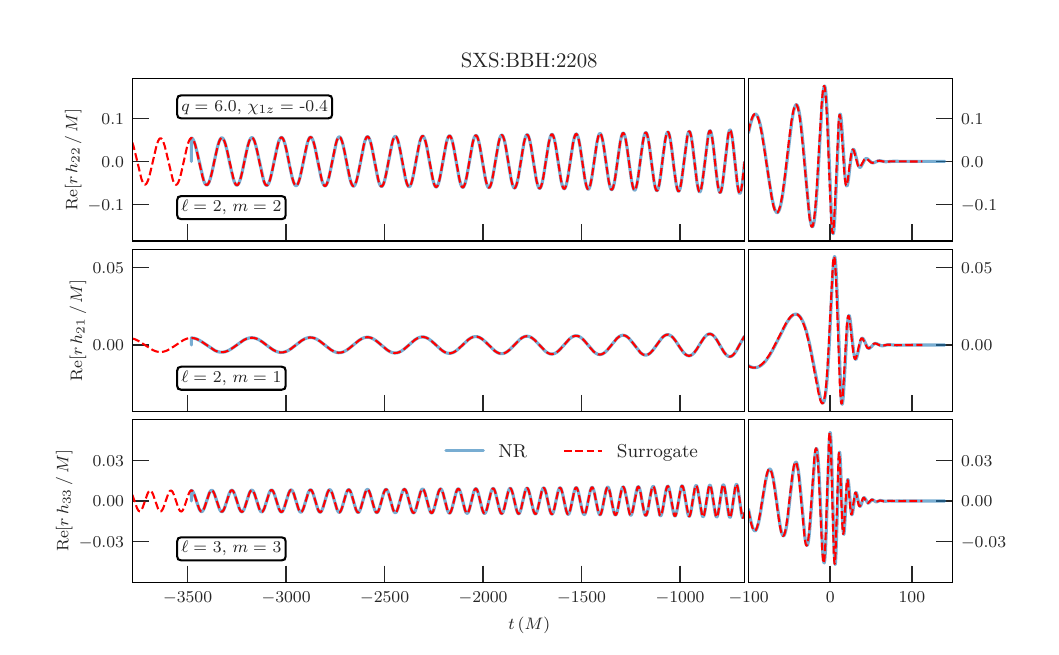}}
    \caption{An example \texttt{WaSABINRHybSur2dq15} waveform. We show the $(2,2)$, $(2,1)$, and $(3,3)$ modes in the top, middle, and bottom rows of the plot. The SXS identifier of the NR waveform is shown at the top of the plot. Portions of the waveforms that appear gray are where the surrogate model and NR become indistinguishable. The surrogate waveform has been shifted in time and phase using the \texttt{gwtools} package to align with the NR waveform, whose peak time has been shifted to $t=0$.}
    \label{fig:sur_NR_comparison_SXS_BBH_2208}
\end{figure*}

In the final step of the surrogate-building procedure, we construct fits over parameter space at each empirical time for each data piece, enabling us to predict the data piece for arbitrary parameter values $\boldsymbol{\lambda}$. To construct these fits, we use the Gaussian process regression (GPR) \cite{10.7551/mitpress/3206.001.0001} fitting method as implemented in the Python package \texttt{scikit-learn} \cite{scikit-learn}, following the procedure outlined in the supplementary material of Ref. \cite{Varma:2018aht} and in Ref. \cite{Taylor:2020bmj}. We parameterize the fits with $\boldsymbol{\lambda} = (\text{log}(q),\, \hat{\chi})$, where
\begin{align}
    \hat{\chi} = \frac{\chi_{\text{eff}} - 38 \eta (\chi_{1z} + \chi_{2z}) / 113}{1 - 76 \eta / 113}
    \label{eq:chi_hat}
\end{align}
is the spin parameter entering the GW phase at leading order in the PN expansion \cite{Khan:2015jqa, Ajith:2011ec, Cutler:1994ys, Poisson:1995ef}, and $\chi_{\text{eff}} = \frac{q \chi_{1z} + \chi_{2z}}{1+q}$ is the effective spin; since we do not model the secondary spin, we set $\chi_{2z}=0$ above. We choose this parameterization for the fits to be consistent with Ref. \cite{Varma:2018mmi}; we note, though, that the fit accuracy, and thereby the model accuracy, improves considerably when using the $\text{log}(q)$ parameterization compared to $q$, in agreement with Refs. \cite{Varma:2018mmi, Yoo:2022erv, Yoo:2023spi, Islam:2022laz, Rink:2024swg}.


By inserting the basis functions, empirical times and parametric fits from the previous steps into Eqs. (18) and (19) of Ref. \cite{Field:2013cfa}, we construct empirical interpolants—denoted by $A_{\ell m}^S$ and $\phi_{\ell m}^S$, where the superscript $S$ stands for ``surrogate"—for the waveform data pieces. During the evaluation stage, only the fitting functions for a given data piece need to be evaluated at its associated nodes once a parameter $\boldsymbol{\lambda}$ is specified by the user, from which time samples the full time series of the data piece can be generated through its empirical interpolant representation; and if the surrogate waveform is required to have a target temporal grid, we interpolate the surrogates for the waveform data pieces from the downsampled grid of Sec. \ref{subsubsec:downsample} to the required grid using cubic splines. Finally, using Eq. (\ref{eq:hlm}), we combine the interpolated waveform data pieces to obtain the surrogate prediction for the waveform modes of the inertial frame strain, which is written as
\begin{align}
    h_{\ell m}^S(t;\boldsymbol{\lambda}) = A_{\ell m}^S(t;\boldsymbol{\lambda})\,e^{-i\,\phi_{\ell m}^S(t;\boldsymbol{\lambda})}. 
\end{align}
In Fig. \ref{fig:sur_NR_comparison_SXS_BBH_2208}, we show waveforms from the \texttt{WaSABINRHybSur2dq15} model and the NR dataset at the binary parameter point $(q,\, \chi_{1z}) = (6,\, -0.4)$.


\begin{table}[h!]
    \centering
    \resizebox{\columnwidth}{!}{%
        \begin{tabular}{c c} 
            \hline \hline
            Data piece & Number of basis functions \\ [1ex] 
            \hline
            $\phi_{33}$ & 16 \\
            $A_{22}$, $\phi_{22}$ & 15 \\ 
            $A_{21}$, $\phi_{21}$, $A_{33}$ & 12 \\
            $A_{44}$, $\phi_{44}$, $A_{55}$, $\phi_{55}$ & 10 \\ [1ex] 
            \hline \hline
        \end{tabular}%
    }
    \caption{Number of basis functions for each waveform data piece.}
    \label{table:basis_functions}
\end{table}

\section{Error Analysis}
\label{sec:error_analysis}
In this section, we examine the quality of \texttt{WaSABINRHybSur2dq15} by assessing the errors in our training data incurred through the hybridization procedure; by using cross-validation methods to evaluate the errors in the surrogate model; and by estimating the efficacy of the model to extrapolate beyond the bounds of its training region.

\subsection{Hybridization Errors}
\label{subsec:hybrid_errors}
In constructing the surrogate training data, we inevitably introduce errors—arising from our choices of hybridization method and inspiral model—that need to be estimated in order to determine the quality of our hybrid waveforms. 

To estimate the errors in each hybrid waveform, we make use of two methods based on those used in Ref. \cite{Yoo:2023spi} for the same purpose. First, we compare the hybrid waveform to the NR waveform it comprises in the matching window. This method allows us to ascertain the errors introduced from the difference between the NR and \texttt{WaSABI-C} waveforms; from using the amplitude-frequency stitching method discussed in Ref. \cite{Varma:2018mmi}; and from using the smooth transition function in Eq. (28) of Ref. \cite{Varma:2018mmi} to stitch the waveforms together within the matching window. Second, to estimate the error introduced from using \texttt{WaSABI-C} waveforms for the early inspiral parts of our hybrids, we compare the hybrid to an NR simulation that has identical binary parameters, but with a much longer initial binary separation. And for these extensive NR simulations, we include an estimate of the resolution error between the two highest-resolution NR waveforms in the matching window. Of the NR simulations we include in the training dataset, 12 of them have simulations that are extensive enough to compute the inspiral error, and contain lower-resolution waveforms.

To compare two waveforms labeled $h_a$ and $h_b$, where $a$ and $b$ are labels and not Einstein indices, we use the following cost function $\mathcal{E}$, defined as the normalized $L^2$ difference between the waveforms in a time segment,
\begin{align}
    \mathcal{E}[h_a, h_b] = \frac{1}{2} \frac{\sum_{\ell,m} \,\int_{t_1}^{t_2}\,|h_{a,\ell m} (t) - h_{b, \ell m}(t)|^2\,dt}{\sum_{\ell, m}\,\int_{t_1}^{t_2}\, |h_{a, \ell m} (t)|^2\,dt},
    \label{eq:L2-norm}
\end{align}
where $t_1$ and $t_2$ indicate the start and end of the time segment. This cost function has been shown to reduce to the weighted average of the time-domain mismatch over the sky (see Appendix C of Ref. \cite{Blackman:2017dfb}).

\begin{figure}
    \centering
    \includegraphics[width=\linewidth]{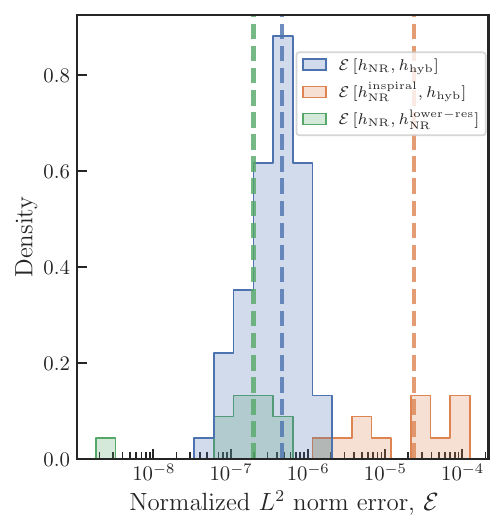}
    \caption{Estimates of the hybridization errors. The measure $\mathcal{E}[h_{\text{NR}}, h_{\text{hyb}}]$, using Eq. (\ref{eq:L2-norm}), computes the error between the NR and hybrid waveforms in the matching window. The measure $\mathcal{E}[h_{\text{NR}}^{\text{inspiral}}, h_{\text{hyb}}]$ computes the error between the NR and hybrid waveforms in a time segment with an end time of $t_2$, the end of the matching window, and an initial time of $t_1=t_2-3000\,M$. We also include $\mathcal{E}[h_{\text{NR}}, h_{\text{NR}}^{\text{lower-res}}]$, the resolution error comparing the highest- and second-highest-resolution NR waveforms within the matching window. While $\mathcal{E}[h_{\text{NR}}, h_{\text{hyb}}]$ was computed for each of the 65 waveforms, $\mathcal{E}[h_{\text{NR}}^{\text{inspiral}}, h_{\text{hyb}}]$ and $\mathcal{E}[h_{\text{NR}}, h_{\text{NR}}^{\text{lower-res}}]$ were computed for the 12 waveforms for which more-extensive and lower-resolution simulations were available. The histograms are normalized such that the area under each curve when integrated over $\text{log}_{10}(\mathcal{E})$ is 1. The dashed vertical lines represent the median values.}
    \label{fig:hybrid-errors-histogram}
\end{figure}

For the first comparison, we compute the error within the matching window between each pair of hybrid and NR waveforms using Eq. (\ref{eq:L2-norm}), where $t_1$ and $t_2$ are the start and end times of the window, respectively. For the second comparison, we first transform the extensive NR waveforms to the frame of their corresponding hybrid waveform, and then we compute the error between these waveforms in a time segment where $t_2$ is the end of the matching window and $t_1 = t_2 - 3000 \, M$. Also, we transform the lower-resolution NR waveforms to the final frame of the extensive NR waveforms, and compute the error between them in the matching window. The results of these comparisons are shown in Fig. \ref{fig:hybrid-errors-histogram}. We find that our estimates for the hybridization and inspiral errors are low, though they tend to be higher than the NR resolution errors. In particular, the inspiral error is highest overall, indicating that a limiting source of error in the training waveforms comes from the \texttt{WaSABI-C} inspiral model. We note, however, that the majority of the extensive NR simulations used to compute the resolution error represent binaries with close-to-comparable masses; considering the difficulty in simulating BBH systems for higher\text{--}mass ratios, we expect the NR resolution error that we computed to underestimate the true resolution error.

\subsection{Surrogate errors}
\label{subsec:surrogate_errors}
\begin{figure}
    \centering
    \includegraphics[width=\linewidth]{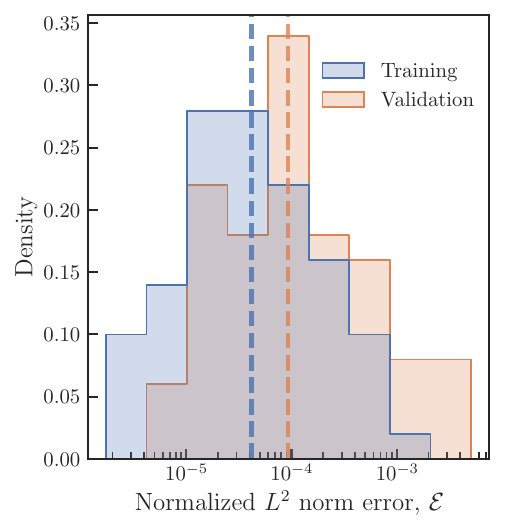}
    \caption{Estimates of the leave-one-out cross-validation errors using all available modes. The training (blue) and validation (red) errors are computed using Eq. (\ref{eq:L2-norm}) as the performance metric. Before the errors are computed, the surrogate and hybrid waveforms are aligned in time and phase using the \texttt{gwtools} package. The histograms are normalized such that the area under each curve when integrated over $\text{log}_{10}(\mathcal{E})$ is 1. The dashed vertical lines represent the median values.}
    \label{fig:training-vs-validation-errors}
\end{figure}

\begin{figure*}
    \centering
    \makebox[\textwidth][c]{\includegraphics[width=1.1\linewidth]{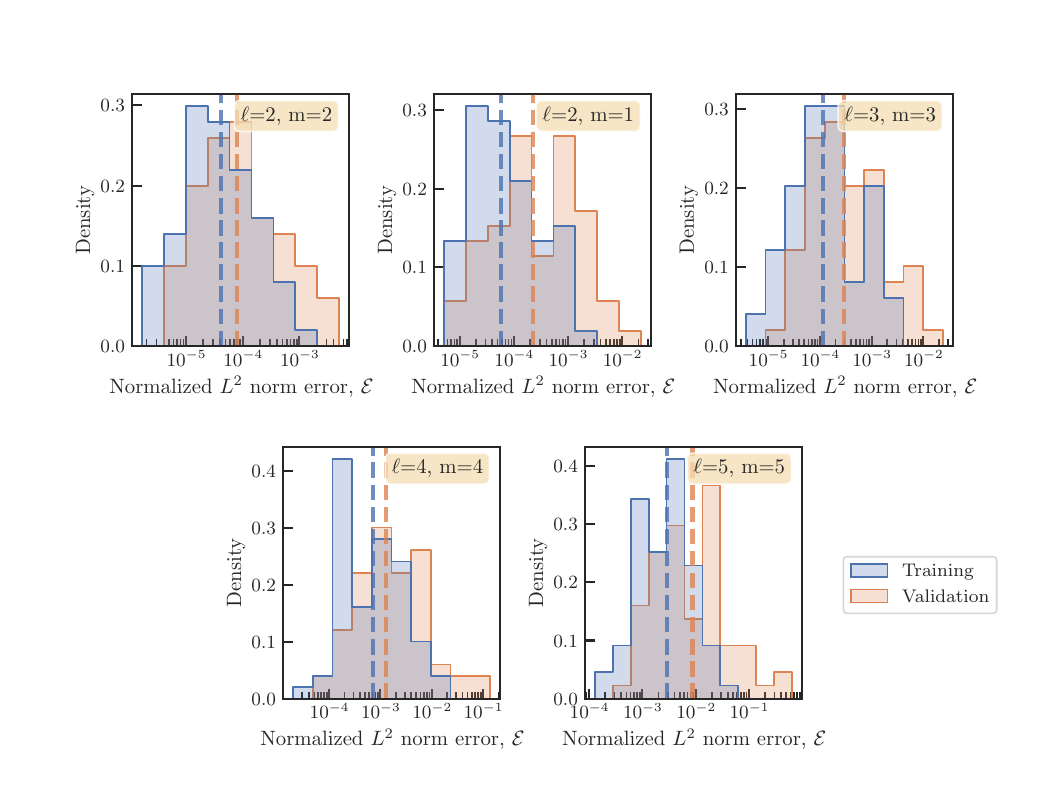}}
    \caption{Estimates of the leave-one-out cross-validation errors for the following individual modes: $(2,2),\, (2,1),\, (3,3),\, (4,4),$ and $(5,5)$. The training (blue) and validation (red) errors are computed using Eq. (\ref{eq:L2-norm}) as the performance metric. Before the errors are computed, the surrogate and hybrid waveforms are aligned in time and phase using the \texttt{gwtools} package. The histograms are normalized such that the area under each curve when integrated over $\text{log}_{10}(\mathcal{E})$ is 1. The dashed vertical lines represent the median values.}
    \label{fig:training-vs-validation-errors-per-mode}
\end{figure*}

We evaluate the errors contracted from constructing the \texttt{WaSABINRHybSur2dq15} surrogate model by using $k$-fold cross-validation, a measure of a model's predictive performance which quantifies how well the model can predict data not used during its training \cite{gelman2013bayesian}. The hyperparameter $k$ indicates the number of folds, or train-test splits—the partition of a dataset into a training set with which to build a model, and a test, or validation, set to which predictions by the model are compared. 




In $k$-fold cross-validation, we randomly divide a dataset $d$ into $k$ disjoint subsets denoted by $d^{(i)}$, where $i$ runs from $1$ to $k$ and also labels the fold. The subsets $d^{(i)}$ are called the validation datasets and will be used to compute the validation errors, a measure of the model's out-of-sample predictive accuracy. For each fold $i$, we put together a training dataset defined as $d^{(\neg i)} = \cup_{j \neq i}\,d^{(j)}$, omitting the validation dataset $d^{(i)}$; we use the training dataset $d^{(\neg i)}$ to construct a trial surrogate model $h^{S_i}$; then, for each sample $d_n^{(i)}$ in $d^{(i)}$ with input parameter vector $\boldsymbol{\lambda_n}$, we evaluate the prediction $h^{S_i}(\boldsymbol{\lambda_n})$ and compute the validation errors defined as $\mathcal{C}[h^{S_i}(\boldsymbol{\lambda_n}),\, d_n^{(i)}]$, where $\mathcal{C}$ is some performance metric. The training errors, a measure of the model's in-sample predictive accuracy, are similarly calculated using $\mathcal{C}[h^{S}(\boldsymbol{\lambda_n}),\, d_n^{(i)}]$, where $h^S$ is the surrogate model constructed from the full dataset $d$. We repeat this procedure $k$ times, iterating through the folds.

\subsubsection{Time-domain errors}
\label{subsubsec:time-domain-errors}
We begin our evaluation of \texttt{WaSABINRHybSur2dq15} with leave-one-out cross-validation (LOO-CV), the special case in which $k=N$, where $N=65$ is the total number of samples in our dataset; each validation set therefore represents a single data point. In short, for each hybrid waveform in our dataset, we train the surrogate model after excluding said hybrid from the training set, and compare the model's prediction of the hybrid to the excluded waveform. To compute the validation errors, we use the normalized $L^2$ norm of Eq. (\ref{eq:L2-norm}) as our performance metric, with $t_1$ and $t_2$ set to the start and end times of the surrogate waveform so that we measure the accuracy of the full waveform. Added to this, we assess the model's ability to faithfully reproduce each sample of the full training set by computing the training errors.

\begin{figure*}
     \centering
     \begin{subfigure}[]{0.475\textwidth}
         \centering
         \includegraphics[width=\textwidth]{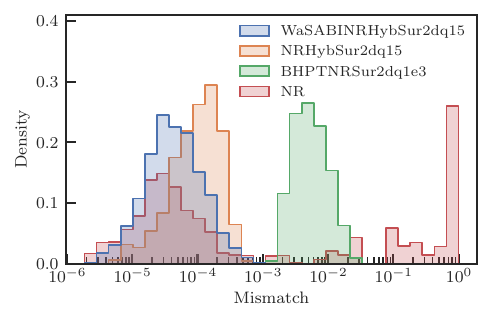}
     \end{subfigure}
     \begin{subfigure}[]{0.495\textwidth}
         \centering
         \includegraphics[width=\textwidth]{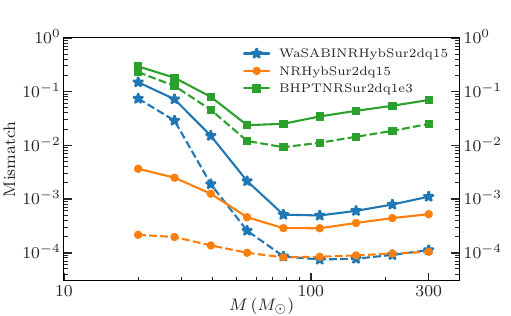}
         \begin{minipage}{0.1cm} 
         \vfill
         \end{minipage}
     \end{subfigure}
        \caption{Frequency-domain mismatches for \texttt{WaSABINRHybSur2dq15}, \texttt{NRHybSur2dq15}, and \texttt{BHPTNRSur2dq1e3} when compared against hybrid waveforms. The mismatches are computed at several points in the sky of the source frame for all available modes in each waveform. For both \texttt{WaSABINRHybSur2dq15} and \texttt{NRHybSur2dq15}, we use the $(2,2),\, (2,1),\, (3,3),\, (4,4),$ and $(5,5)$ modes, whereas for \texttt{BHPTNRSur2dq1e3}, we use $\ell \leq 4$ modes except the $(4,1)$ and $m=0$ modes. Note that the mismatches for \texttt{WaSABINRHybSur2dq15} are out-of-sample errors. \textit{\textbf{Left:}} Mismatches computed using a flat noise curve for the late inspiral part of the waveforms, which have been truncated such that their starting time is $t=-1000\,M$ before the peak, ensuring that we are only comparing the NR part of each waveform. For comparison with the model mismatches, we include the NR resolution error, which is computed by comparing the two highest-available-resolution NR waveforms. The histograms are normalized such that the area under each curve when integrated over $\text{log}_{10}(\mathcal{E})$ is 1. \textit{\textbf{Right:}} Mismatches computed using the Advanced-LIGO noise curve as a function of total mass. The solid (dashed) lines represent the 95th percentile (median) mismatch value.}
        \label{fig:model-comparison-mismatches}
\end{figure*}

We display the LOO-CV results in Figs. \ref{fig:training-vs-validation-errors} and \ref{fig:training-vs-validation-errors-per-mode}. Figure \ref{fig:training-vs-validation-errors} shows the training and validation errors computed using all available modes, whereas Fig. \ref{fig:training-vs-validation-errors-per-mode} presents a mode-by-mode comparison of the training and validation errors. We see that the median of the validation error is less than $10^{-3}$ for waveform modes up to $(4,4)$, increasing to $10^{-2}$ for the case of the $(5,5)$ mode. Across all modes, the histograms of both errors broadly overlap, indicating that the surrogate model can predict the quadrupolar and subdominant modes for an unknown sample without losing considerable accuracy.

\subsubsection{Frequency-domain mismatches}
\label{subsubsec:freq-domain-mismatches}
In addition to the LOO-CV analysis, we evaluate the accuracy of \texttt{WaSABINRHybSur2dq15} by computing mismatches between the surrogate and the hybrid waveforms used to train the model. Specifically, we use $k$-fold cross-validation where $k=13$. With this choice, we construct exactly 13 surrogate models, each of which has 5 hybrid waveforms left out of its training set. For each trial surrogate, we compute mismatches against the hybrid waveforms that were left out, i.e. the validation waveforms, providing us with another estimate of the out-of-sample error. For comparison with the model errors, we also compute mismatches between the hybrid waveforms and two other surrogate models: \texttt{NRHybSur2dq15} \cite{Yoo:2022erv} (valid for $q \leq 15$, $|\chi_{1z}| \leq 0.5$) and \texttt{BHPTNRSur2dq1e3} \cite{Rink:2024swg, Rifat:2019ltp} (valid for $3 \leq q \leq 1000$, $|\chi_{1z}| \leq 0.8$). Whereas \texttt{NRHybSur2dq15} contains the same set of modes as \texttt{WaSABINRHybSur2dq15}, \texttt{BHPTNRSur2dq1e3} includes all $\ell \leq 4$ modes except for the $(4,1)$ and $m=0$ modes.

\begin{figure*}
    \centering
    \makebox[\textwidth][c]{\includegraphics[width=1.1\linewidth]{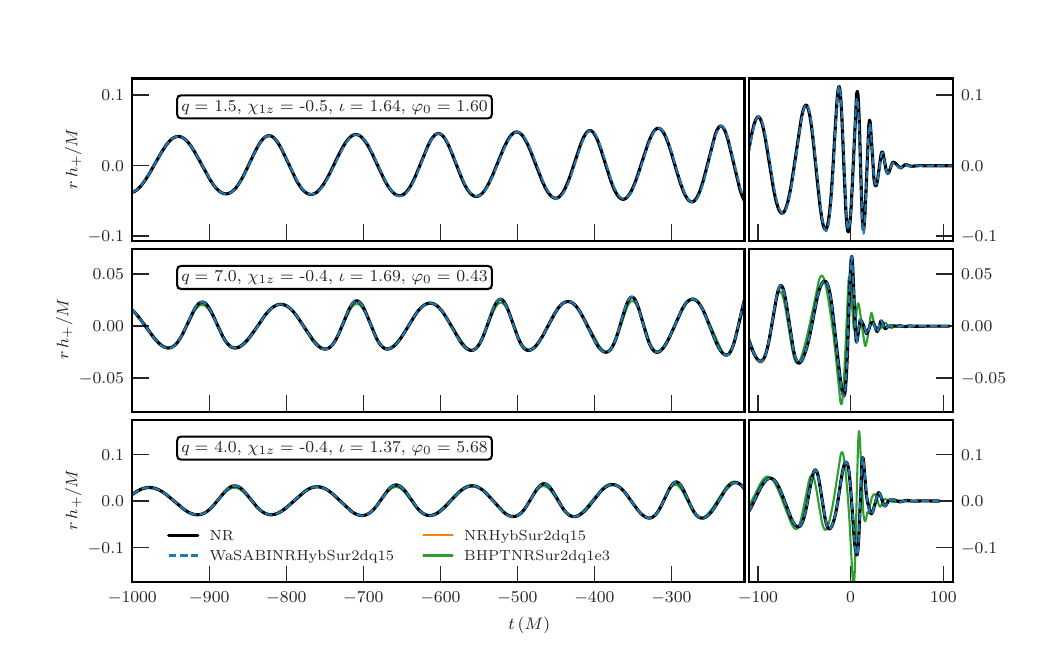}}
    \caption{The plus polarization of the waveforms for the cases resulting in the largest mismatches for \texttt{WaSABINRHybSur2dq15} (top row), \texttt{NRHybSur2dq15} (middle row), and \texttt{BHPTNRSur2dq1e3} (bottom row) in the left panel of Fig. \ref{fig:model-comparison-mismatches}. For comparison with the model waveforms, we include the hybrid waveforms, here labeled as NR because we only show the late inspiral, merger, and ringdown. In each row, the waveforms have been projected along the direction which results in the largest mismatch, using all the available modes of the given model. The binary parameters and direction in the source frame leading to the largest mismatches for \texttt{WaSABINRHybSur2dq15} (top row), \texttt{NRHybSur2dq15} (middle row), and \texttt{BHPTNRSur2dq1e3} (bottom row) are shown in the boxed insets. Note that each \texttt{WaSABINRHybSur2dq15} waveform is evaluated using a trial surrogate that did not include the hybrid waveform in its training dataset. All waveforms have been shifted in time such that their peak amplitudes is located at $t=0$; and they have been rotated about the $\hat{z}$-axis such that their orbital phase is zero at $t=-1000\,M$.}
    \label{fig:worst_mismatches}
\end{figure*}

To estimate the difference between two waveforms $h_a$ and $h_b$, we compute the frequency-domain mismatch $\mathcal{M}$, 
\begin{align}
    \mathcal{M} = 1 - \frac{\langle h_a, h_b \rangle}{\sqrt{\langle h_a, h_a \rangle \langle h_b, h_b \rangle}},
    \label{eq:mismatch}
\end{align}
where $\langle \cdot, \cdot \rangle$ is the frequency domain noise-weighted inner product
\begin{align}
    \langle h_a, h_b \rangle = 4 \text{Re} \left[\int_{f_{\text{min}}}^{f_{\text{max}}}\, \frac{\tilde{h}_a(f)\,\tilde{h}_b^*(f)}{S_n(f)}\,df \right].
    \label{eq:inner_product}
\end{align}
Here, $\tilde{h}(f)$ is the Fourier transform of the complex waveform $h(t)$; the operator $*$ indicates complex conjugation; Re denotes the real part; and $S_n(f)$ is the one-sided power spectral density of noise in a GW detector. Before computing the mismatches, we first taper both ends of the time-domain waveform using a Planck window \cite{McKechan:2010kp}; the tapering at the start of the waveform is done over the first 1.5 GW cycles of the $(2,2)$ mode, and the tapering at the end over the last $20\,M$. Then, we zero-pad the waveform to the nearest power of two.

We optimize the mismatches over shifts in time, polarization angle, and initial orbital phase following the procedure described in Appendix D of Ref. \cite{Blackman:2017dfb}. To treat both plus and cross polarizations on an equal footing, we assume a network of two detectors in which one measures only the plus polarization, and the other only the cross polarization. For each pair of waveforms, we compute the mismatches at 37 points uniformly distributed on the sky in the source frame.

Finally, when computing flat noise $(S_n=1)$ mismatches, we choose $f_{\text{min}}$ to be the frequency of the $(2,2)$ mode at the end of the initial tapering window, and $f_{\text{max}} = 5\,f_{22}^{\text{peak}}$, where $f_{22}^{\text{peak}}$ is the frequency of the $(2,2)$ mode at its peak, ensuring that we capture the peak frequency of every mode considered in this work. We also compute mismatches using the Advanced-LIGO design sensitivity Zero-Detuned-HighP noise curve \cite{aLIGO_noise_curve} with $f_{\text{min}}=20\,\text{Hz}$ and $f_{\text{max}}=2000\,\text{Hz}$.

Figure \ref{fig:model-comparison-mismatches} shows mismatches of \texttt{WaSABINRHybSur2dq15}, \texttt{NRHybSur2dq15}, and \texttt{BHPTNRSur2dq1e3} against the hybrid waveforms. In computing the mismatches, we use all available modes for each waveform model. In the left panel of Fig. \ref{fig:model-comparison-mismatches}, we show mismatches computed using a flat noise curve over the NR part (or late inspiral part) of the hybrid waveforms, which we set up by truncating and starting the initial tapering at $t=-1000\,M$; we choose this truncation time because the end time of the hybridization window for each hybrid waveform is generally located more than $1000\,M$ from the peak of the $L^2$-norm of the strain (we find there are only 5 hybrid waveforms for which the end time of the window is less than $1000\,M$ from the peak). We see that the mismatches of \texttt{WaSABINRHybSur2dq15} are about less than an order of magnitude lower than those of \texttt{NRHybSur2dq15}, whereas the mismatches of \texttt{BHPTNRSur2dq1e3} are larger than the other two surrogate models by about two orders of magnitude. We then compare these mismatches with the truncation error in the NR simulations, which we measure by computing the mismatches between the two highest\text{--}available resolutions of each NR waveform. We find that the errors in \texttt{WaSABINRHybSur2dq15} are well within the NR truncation error. We note, however, that this truncation error is a conservative estimate, for we take the highest resolution NR waveform to be the fiducial waveform in computing the mismatches, leading us to interpret the NR error in Fig. \ref{fig:model-comparison-mismatches} to be the error in the lower-resolution simulation.




In the right panel of Fig. \ref{fig:model-comparison-mismatches}, we show mismatches computed using the Advanced-LIGO noise curve for the surrogate models against the hybrid waveforms. As these mismatches are dependent on total mass, we show the mismatch for total masses $M$ between $20\,M_{\odot}$ and $300\,M_{\odot}$. At each total mass, we show the median and 95th percentile mismatches. In the mass range $80\,M_{\odot} \leq M \leq 300\,M_{\odot}$, \texttt{WaSABINRHybSur2dq15} has 95th percentile mismatches that are about the same order of magnitude as those of \texttt{NRHybSur2dq15}, and at least two orders of magnitude smaller than those of \texttt{BHPTNRSur2dq1e3}, in agreement with the left panel of Fig. \ref{fig:model-comparison-mismatches}. Moreover, the mismatches for \texttt{NRHybSur2dq15} are broadly uniform across all masses. The discrepancy in mismatches between \texttt{WaSABINRHybSur2dq15} and \texttt{NRHybSur2dq15} at low\text{--}total masses can be explained by considering each model's regime of validity for all available modes with respect to total mass: \texttt{WaSABINRHybSur2dq15} is valid for all modes for $M \gtrsim 111.5\,M_{\odot}$ (see Sec. \ref{subsubsec:downsample}); \texttt{NRHybSur2dq15} for $M \gtrsim 9.5\,M_{\odot}$ \cite{Yoo:2022erv}.

We present in Fig. \ref{fig:worst_mismatches} waveforms generated by \texttt{WaSABINRHybSur2dq15}, \texttt{NRHybSur2dq15}, and \texttt{BHPTNRSur2dq1e3} for the instances leading to the largest mismatches in the left panel of Fig. \ref{fig:model-comparison-mismatches}. Even for its worst case, \texttt{WaSABINRHybSur2dq15} shows good agreement with the NR part of the hybrid waveform. We note that we align all the waveforms in time and orbital phase.



\subsection{Extrapolating outside training region}
\label{subsec:extrapolate}
\begin{figure}
    \centering
    \includegraphics[width=\linewidth]{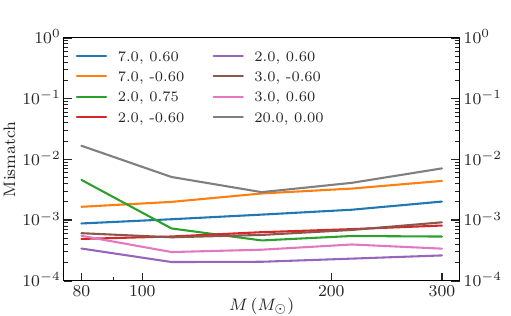}
    \caption{Frequency-domain mismatches between \texttt{WaSABINRHybSur2dq15} and NR simulations with binary parameters beyond the training region (see Fig. \ref{fig:sxs-dataset}). The mismatches shown are computed using the Advanced-LIGO noise curve over a range of total masses at several points in the sky of the source frame. We use the same set of modes in the NR waveform as the surrogate to best assess the accuracy of the extrapolation. The solid lines represent the 95th percentile mismatch values. The numbers in the legend indicate the simulation's mass ratio $q$ and primary spin $\chi_{1z}$.}
    \label{fig:extrapolation}
\end{figure}

Thus far, the errors we have considered have been restricted to the training parameter region of the \texttt{WaSABINRHybSur2dq15} model $(1.25 \leq q \leq 15,\, |\chi_{1z}| \leq 0.5,\, \chi_{2z}=0)$. This does not, however, preclude us from assessing the model's efficacy to extrapolate outside its training region. We perform this assessment by comparing the model against a set of existing simulations (SXS:BBH:0202, 0207, 0615, 2110, 2129, 2134, 2161, and 2516) \cite{SXS, Boyle:2019kee, Scheel:2025jct} that have high primary spin magnitudes $(|\chi_{1z}| > 0.5)$ or mass ratios $(q > 15)$ relative to the training set—represented by triangle markers in Fig. \ref{fig:sxs-dataset}—the mismatch results of which we present in Fig. \ref{fig:extrapolation}. Note that we do not extrapolate below $q=1.25$, as we only test points which lie in the regimes of validity of both \texttt{WaSABI-C} and NR.

We find that the surrogate model's extrapolative accuracy is largely dependent on the region of parameter space in which the model is evaluated, as evinced by the wide range of mismatches across the test cases in Fig. \ref{fig:extrapolation}. For simulations with close-to-comparable mass ratios, the mismatches range from $2\times 10^{-4}$ to $10^{-3}$ in the model's regime of validity $(M \gtrsim 111.5\,M_{\odot})$. Contrastingly, mismatches for the $q=7$, $|\chi_{1z}|=0.6$ cases lie above $10^{-3}$, with $(q,\, \chi_{1z}) = (7,\,-0.6)$ faring better than $(q,\, \chi_{1z}) = (7,\,0.6)$ over the entire range of total masses, the former reaching mismatches of about $2\times 10^{-3}$ at higher total masses compared to the latter's $5\times 10^{-3}$. These mismatches can be explained by inspecting the parameter space of the training set shown in Fig. \ref{fig:sxs-dataset}: near the low\text{--}mass ratio boundary, the training space is more populated with simulations along the upper\text{--} and lower\text{--}primary spin bounds compared to the region around $q=7$, which contains a single point on the lower-$\chi_{1z}$ boundary at $q=8$ and none on the upper boundary; the lack of boundary points along the primary-spin dimension around $q=8$ suggests that the model's parametric fits lack sufficient data in these regions, and so the model does worse here than in the low\text{--}mass ratio regime. Lastly, the highest mismatch result corresponds to the simulation with the most extreme parameters considered in this analysis, $(q,\, \chi_{1z}) = (20,\,0)$, a point far outside the training region, and at which we expect the model's accuracy to diminish.

Given that many of the mismatches lie below $10^{-3}$ (cf. \texttt{WaSABINRHybSur2dq15} errors in the right panel of Fig. \ref{fig:model-comparison-mismatches}); and given the difficulty in assessing the accuracy of the model for $15 < q < 20$, owing to the lack of NR simulations in this region of parameter space; we therefore recommend the surrogate be used for $1.25 \leq q \leq 15$ and $|\chi_{1z}| \leq 0.6$, and advise caution when extrapolating the model.



\section{Conclusions}
\label{sec:conclusions}
In this work, we present \texttt{WaSABINRHybSur2dq15}, a new surrogate waveform model aimed at non-precessing, asymmetric-mass BBHs with zero secondary spin. The model is the first to combine SF, PN, and NR data, and is trained on 65 BBH hybrid waveforms, each of which is constructed by stitching together \texttt{WaSABI-C} inspirals and SXS NR waveforms, whose binary parameters lie within the domain $1.25 \leq q \leq 15$ and $|\chi_{1z}| \leq 0.5$, all other spins having been set to zero; and whose spin-weighted spherical harmonic modes include $(2,2),\, (2,1),\, (3,3),\,(4,4)$, and $(5,5)$. The model spans the entire LVK frequency band (with $f_{low}=20$ Hz) for total masses $M \gtrsim 111.5\,M_{\odot}$. And as more NR simulations become available, the accuracy of the surrogate model can be improved by adding newer hybrid waveforms without removing any of the previous ones, owing to the model's hierarchical structure.

By conducting a $k$-fold cross-validation study, where $k=13$, we find that \texttt{WaSABINRHybSur2dq15} can faithfully reproduce the hybrid waveforms used to train the model with median mismatches of about $10^{-4}$ and 95th percentile mismatches less than $10^{-3}$ for total masses in the range $80\,M_{\odot} \leq M \leq 300\,M_{\odot}$. These errors are comparable to those of the aligned-spin surrogate model \texttt{NRHybSur2dq15}. The model will soon be made publicly available as part of the \texttt{GWSurrogate} package \cite{Field:2025isp}.

Though the self-force waveform models in the \textrm{W\footnotesize A\normalsize SABI} package are currently restricted to quasi-circular inspirals with aligned-spins or precessing secondary spin, work is ongoing to extend these models to incorporate more dynamics of binaries, such as eccentricity, generic precession of both spins, and second-order SF results in Kerr, which, when combined with the latest NR results, will enable us to develop more comprehensive surrogate models, covering vaster portions of the parameter space.

\acknowledgements
HI thanks Josh Mathews for introducing him to the WaSABI package in the Black Hole Perturbation Toolkit; Geoffrey Comp\`ere for bringing the \texttt{WaSABI-C} model to his attention; and Scott Field for giving helpful advice with regard to integrating the \texttt{WaSABINRHybSur2dq15} model into the \texttt{GWSurrogate} package. HI additionally thanks Aasim Jan for sharing his mismatch code with him. Mismatch calculations were performed on the CIT cluster of the LIGO Laboratory, which is supported by NSF Grants PHY-0757058 and PHY-0823459. This research was supported by NASA grant No. 80NSSC24K0437, NSF PHY-2207780, and ERC grant EMRIWaveforms (DOI: \href{https://doi.org/10.3030/101200625}{10.3030/101200625}). This work makes use of public NR data from the SXS Collaboration \cite{Scheel:2025jct}; the Black Hole Perturbation Toolkit \cite{BHPToolkit}; Python libraries including \texttt{SciPy} \cite{Virtanen:2019joe}, \texttt{NumPy} \cite{Harris:2020xlr}, \texttt{scikit-learn} \cite{scikit-learn}, \texttt{matplotlib} \cite{Hunter:2007}, and \texttt{seaborn} \cite{Waskom2021}; and the \texttt{gwsurrogate} \cite{Field:2025isp, gwsurrogatepypi}, \texttt{gwtools} \cite{gwtoolspypi}, and \texttt{RomPy} \cite{rompypypi} packages. Furthermore, this work was done by members of the Weinberg Institute and has an identifier of WI Preprint UT-WI-19-2026.

\appendix
\section{NR Simulations}
\label{appendix:NR}

We use the following dataset from the third SXS public catalog (Table \ref{table:sim_list}) to hybridize NR waveforms with \texttt{WaSABI-C} inspirals \cite{Honet:2025gge, Honet:2025lmk}. These simulations were chosen because (i) their mass ratios span from $q=1.25$ to $q=15$, (ii) their dimensionless primary spins along the orbital angular momentum are less than or equal to 0.5, and (iii) they have zero secondary spin.

\begin{table*}
\begin{minipage}[t]{\columnwidth}
    \centering
    \begin{tabular}[t]{c c c c c c }
      \hline \hline \\ 
SXS ID  & $q$ & $\chi_{1z}$ & $M \omega_{\text{ref}}$ & $N_{\text{orbits}}$ & $e_{\text{ref}}$ \\ [0.5ex] 
\hline  
SXS:BBH:1143 & 1.25 & -1.367e-04 & 2.229e-02 & 9.11 & 8.760e-05 \\
SXS:BBH:3706 & 1.41 & 1.987e-02 & 1.733e-02 & 15.01 & 2.155e-04 \\
SXS:BBH:2335 & 1.5 & -5.000e-01 & 1.642e-02 & 14.8 & 4.800e-05 \\
SXS:BBH:2332 & 1.5 & 5.000e-01 & 1.975e-02 & 13.43 & 1.493e-04 \\
SXS:BBH:0593 & 1.5 & 3.198e-05 & 1.663e-02 & 16.31 & 5.750e-05 \\
SXS:BBH:1354 & 1.83 & -1.530e-04 & 1.703e-02 & 16.21 & 5.310e-05 \\
SXS:BBH:0599 & 2 & 1.999e-01 & 1.721e-02 & 17.2 & 6.795e-04 \\
SXS:BBH:2425 & 2 & -7.636e-05 & 1.938e-02 & 12.91 & 2.490e-04 \\
SXS:BBH:2115 & 2 & -3.000e-01 & 1.509e-02 & 19.32 & 7.620e-05 \\
SXS:BBH:2124 & 2 & 3.000e-01 & 1.639e-02 & 19.3 & 9.610e-05 \\
SXS:BBH:0201 & 2.32 & 6.507e-05 & 1.871e-02 & 14.48 & 1.938e-04 \\
SXS:BBH:2360 & 3 & -4.999e-01 & 2.050e-02 & 10.5 & 2.229e-04 \\
SXS:BBH:1178 & 3 & -1.483e-05 & 2.125e-02 & 12.26 & 1.395e-04 \\
SXS:BBH:2143 & 3 & -3.001e-01 & 1.656e-02 & 18.16 & 3.010e-05 \\
SXS:BBH:2154 & 3 & 3.000e-01 & 1.709e-02 & 20.87 & 7.090e-05 \\
SXS:BBH:2362 & 3 & 4.999e-01 & 2.501e-02 & 11.15 & 5.950e-05 \\
SXS:BBH:2483 & 3.5 & 3.744e-05 & 1.571e-02 & 23.84 & 2.336e-04 \\
SXS:BBH:3631 & 4 & 5.088e-05 & 2.118e-02 & 13.98 & 2.058e-04 \\
SXS:BBH:1937 & 4 & 4.000e-01 & 2.002e-02 & 18.63 & 3.934e-04 \\
SXS:BBH:1961 & 4 & -4.001e-01 & 1.817e-02 & 16.16 & 3.356e-04 \\
SXS:BBH:2484 & 4.5 & 1.598e-05 & 1.644e-02 & 24.82 & 8.098e-04 \\
SXS:BBH:2385 & 5 & 4.999e-01 & 2.096e-02 & 20.4 & 1.680e-04 \\
SXS:BBH:2374 & 5 & -8.411e-05 & 1.795e-02 & 22.06 & 3.260e-05 \\
SXS:BBH:2164 & 6 & 1.439e-06 & 2.108e-02 & 17.53 & 4.762e-04 \\
SXS:BBH:2185 & 6 & 4.000e-01 & 2.181e-02 & 20.2 & 5.958e-04 \\
SXS:BBH:2208 & 6 & -3.999e-01 & 2.017e-02 & 15.55 & 2.217e-04 \\
SXS:BBH:2488 & 6.5 & 4.810e-05 & 1.653e-02 & 30.71 & 7.904e-04 \\
SXS:BBH:0192 & 6.58 & 2.886e-05 & 2.168e-02 & 17.49 & 4.350e-05 \\
SXS:BBH:2491 & 7 & 2.744e-05 & 1.650e-02 & 32.35 & 3.607e-04 \\
SXS:BBH:0205 & 7 & -3.999e-01 & 1.344e-02 & 42.36 & 6.330e-05 \\
SXS:BBH:0203 & 7 & 3.999e-01 & 1.314e-02 & 57.9 & 1.370e-05 \\
SXS:BBH:0188 & 7.19 & 4.612e-05 & 2.344e-02 & 15.54 & 1.504e-04 \\
SXS:BBH:2490 & 7.5 & -5.541e-05 & 1.647e-02 & 34.02 & 5.635e-04 \\ [1ex]


\hline 
\hline

    \end{tabular}
\end{minipage}\hfill 
\begin{minipage}[t]{\columnwidth}
    \centering
    \begin{tabular}[t]{c c c c c c }
      \hline \hline \\ 
SXS ID  & $q$ & $\chi_{1z}$ & $M \omega_{\text{ref}}$ & $N_{\text{orbits}}$ & $e_{\text{ref}}$ \\ [0.5ex] 
\hline  
SXS:BBH:0195 & 7.76 & 2.557e-05 & 2.500e-02 & 14.1 & 2.061e-04 \\
SXS:BBH:2680 & 8 & 4.001e-01 & 1.777e-02 & 36.44 & 6.513e-04 \\
SXS:BBH:2785 & 8 & -4.000e-01 & 1.651e-02 & 29.54 & 6.801e-04 \\
SXS:BBH:2707 & 8 & -1.919e-05 & 1.711e-02 & 32.81 & 3.110e-04 \\
SXS:BBH:3623 & 8 & -4.998e-01 & 2.131e-02 & 15.04 & 2.466e-04 \\
SXS:BBH:0186 & 8.27 & 3.197e-06 & 2.063e-02 & 22.59 & 1.529e-04 \\
SXS:BBH:2495 & 9 & 3.135e-05 & 1.636e-02 & 39.2 & 1.877e-04 \\
SXS:BBH:0189 & 9.17 & 1.183e-05 & 2.069e-02 & 24.13 & 1.020e-04 \\
SXS:BBH:2494 & 9.5 & -1.774e-05 & 1.590e-02 & 43.32 & 1.652e-04 \\
SXS:BBH:0185 & 10 & 1.281e-05 & 2.142e-02 & 23.72 & 2.772e-04 \\
SXS:BBH:2469 & 10 & 5.000e-01 & 2.313e-02 & 27.51 & 2.095e-04 \\
SXS:BBH:2470 & 10.58 & 4.193e-01 & 2.209e-02 & 29.93 & 2.178e-04 \\
SXS:BBH:2466 & 11.22 & -3.622e-01 & 2.077e-02 & 21.94 & 1.653e-04 \\
SXS:BBH:2467 & 11.5 & 5.000e-01 & 2.401e-02 & 28.55 & 5.868e-04 \\
SXS:BBH:2472 & 11.82 & -1.802e-02 & 2.268e-02 & 23.26 & 2.227e-04 \\
SXS:BBH:4260 & 12 & -5.000e-01 & 1.692e-02 & 35.17 & 3.722e-04 \\
SXS:BBH:4236 & 12 & 4.999e-01 & 1.932e-02 & 44.74 & 3.823e-04 \\
SXS:BBH:2474 & 12.94 & -5.000e-01 & 2.135e-02 & 20.46 & 5.306e-04 \\
SXS:BBH:2475 & 13.04 & -3.119e-01 & 2.171e-02 & 22.6 & 2.785e-04 \\
SXS:BBH:2476 & 13.2 & 5.000e-01 & 2.477e-02 & 29.95 & 2.620e-04 \\
SXS:BBH:2479 & 14 & -2.500e-01 & 2.305e-02 & 21.4 & 5.098e-04 \\
SXS:BBH:2481 & 14 & 2.500e-01 & 2.348e-02 & 29.44 & 3.372e-04 \\
SXS:BBH:2480 & 14 & 7.625e-06 & 2.292e-02 & 26.14 & 3.225e-04 \\
SXS:BBH:2482 & 14 & 5.000e-01 & 2.438e-02 & 32.37 & 7.455e-04 \\
SXS:BBH:2478 & 14 & -5.000e-01 & 2.257e-02 & 18.55 & 1.370e-04 \\
SXS:BBH:2465 & 14.91 & 3.132e-01 & 2.431e-02 & 30.11 & 1.699e-04 \\
SXS:BBH:2473 & 15 & -3.518e-01 & 2.238e-02 & 22.47 & 1.020e-04 \\
SXS:BBH:2463 & 15 & -5.000e-01 & 2.287e-02 & 18.76 & 2.962e-04 \\
SXS:BBH:2471 & 15 & -1.630e-01 & 2.368e-02 & 22.48 & 2.976e-04 \\
SXS:BBH:2468 & 15 & 6.457e-02 & 2.325e-02 & 27.95 & 1.552e-04 \\
SXS:BBH:2464 & 15 & 5.000e-01 & 2.544e-02 & 31.55 & 2.666e-04 \\
SXS:BBH:2477 & 15 & 1.398e-05 & 2.413e-02 & 24.36 & 3.746e-04 \\ [1ex] 

\hline 
\hline
    \end{tabular}
  \end{minipage}
  \caption{Configurations for the 65 NR simulations from the third SXS catalog used to construct hybrids to train the surrogate model. Column 1 shows the simulation ID in the catalog. Column 2 shows the mass ratio of the simulation at the reference time. Column 3 shows the dimensionless spin of the primary BH along the direction of the orbital angular momentum at the reference time. Column 4 shows the magnitude of the orbital angular frequency at the reference time. Column 5 shows the number of orbits until the formation of a common apparent horizon, and column 6 shows the eccentricity at the reference time.} 
\label{table:sim_list} 
\end{table*}

\bibliographystyle{ieeetr}
\bibliography{references}

\end{document}